\pdfoutput=1
\documentclass[10pt,twocolumn,letterpaper]{article}

\usepackage[letterpaper,top=0.78in,bottom=0.82in,left=0.68in,right=0.68in,columnsep=0.26in]{geometry}
\usepackage{mathptmx}
\usepackage{amsmath,amssymb,amsthm,mathtools}
\usepackage{graphicx}
\usepackage{booktabs,tabularx}
\usepackage{xcolor}
\usepackage{caption}
\usepackage{fancyhdr}
\usepackage{microtype}
\usepackage[numbers,sort&compress]{natbib}
\usepackage{hyperref}

\definecolor{apsblue}{RGB}{0,90,156}
\hypersetup{
  colorlinks=true,
  linkcolor=apsblue,
  citecolor=apsblue,
  urlcolor=apsblue,
  pdftitle={Penalty-Free Constrained Optimization on Dynamic Quantum Circuits via Phase-Protected Tunneling and Closed-Loop Spectral Ratchets},
  pdfauthor={Chia-Ho Ou}
}

\renewcommand{\thesection}{\Roman{section}}

\makeatletter
\renewcommand{\section}{\@startsection{section}{1}{\z@}%
  {-2.2ex \@plus -0.6ex \@minus -0.2ex}%
  {1.1ex \@plus 0.2ex}%
  {\normalfont\normalsize\bfseries\centering\MakeUppercase}}
\renewcommand{\subsection}{\@startsection{subsection}{2}{\z@}%
  {-1.6ex \@plus -0.4ex \@minus -0.2ex}%
  {0.7ex \@plus 0.2ex}%
  {\normalfont\small\bfseries}}
\makeatother

\newtheorem{theorem}{Theorem}
\newtheorem{proposition}{Proposition}

\begin{document}

\twocolumn[
  \begin{@twocolumnfalse}
    \begin{center}
      {\Large\bfseries Penalty-Free Constrained Optimization on Dynamic Quantum Circuits via Phase-Protected Tunneling and Closed-Loop Spectral Ratchets}\par\vspace{9pt}
      {\normalsize Chia-Ho Ou$^{1,2,*}$}\par\vspace{4pt}
      {\small\itshape
        $^{1}$Department of Computer Science and Information Engineering, National Pingtung University, Pingtung 900391, Taiwan\par
        $^{2}$Graduate School of Information Sciences, Tohoku University, Sendai 980-8579, Japan
      }\par\vspace{3pt}
      {\footnotesize $^{*}$Electronic address: \href{mailto:cho@mail.nptu.edu.tw}{cho@mail.nptu.edu.tw}}\par\vspace{9pt}
    \end{center}
    \begin{quote}\small
      \noindent\textbf{Abstract.} Combinatorial optimization problems subject to hard operational constraints arise across numerous physical and engineering systems. However, existing gate-model quantum optimization schemes generally embed hard constraints into an unconstrained Hamiltonian via large scalar penalty multipliers, and thus induce quadratic spectral aliasing during phase estimation and increase two-qubit gate depth. Accordingly, this paper proposes a penalty-free, closed-loop dynamic quantum circuit scheme, designated as Zeno-decoupled gauge-ratchet optimization (Z-GRO), for solving constrained binary optimization problems without outer-loop parameter training. In the proposed approach, reversible feasibility phase operators interleaved within symmetric dynamical-decoupling echoes confine state evolution to the feasible subspace, while an interaction-correlated commutator gauge operator resolves the initial symmetry stall. Directional cooling is achieved by coupling the unpenalized objective Hamiltonian to a reusable ancilla with a zero-two-qubit-gate threshold phase offset, while conditional feed-forward bypasses echoed tunneling gates on retained low-energy branches. A survival-filtered Ramsey quadrature controller advances the threshold upward from a certified linear-programming lower bound with automatic overshoot rollback. Exact density-matrix and matrix-free quantum trajectory simulations on frustrated spin-glass and quadratic knapsack ensembles show that Z-GRO achieves a single-check discrimination gain exceeding four orders of magnitude. Furthermore, Z-GRO improves the optimal ground-state probability by factors of 3.0 to 4.3 over fixed-schedule quantum Zeno dynamics and 11.4 to 22.5 over penalized variational and feedback schemes. Moreover, Z-GRO reduces the exponential time-to-solution scaling base to 1.529 on the spin-glass ensemble (compared to 1.684 for quantum Zeno dynamics and 1.909 to 2.047 for penalized schemes) and lowers the 14-qubit time-to-solution by factors of 2.4 to 14.5 on the quadratic knapsack ensemble. Under two-qubit gate error rates between $5 \times 10^{-4}$ and $5 \times 10^{-3}$, Z-GRO maintains an optimal probability 2.9 to 7.0 times that of the noisy penalized feedback scheme at a fixed depth, although the ancilla check cost under dense coupling prevents a net ratchet gain beyond the first cycle. In addition, the majority-shielded ratchet limits the loss in the optimal probability to 6.3\% at a 5.0\% mid-circuit readout bit-flip rate. The threshold controller records 0/24, 0/25, and 0/96 overshoots on the noiseless 8-qubit, spin-glass scaling, and readout-noise suites, alongside 3/25 on knapsack scaling and 2/80 late-cycle overshoots under gate noise.
    \end{quote}
    \vspace{10pt}
    \hrule
    \vspace{14pt}
  \end{@twocolumnfalse}
]

\section{Introduction}

Recent advancements in quantum information processing and low-latency cryogenic and room-temperature control electronics have facilitated the development of dynamic quantum circuits for a wide variety of real-world optimization problems, including semiconductor wafer-transfer scheduling, multi-antenna wireless resource allocation, electric power-grid unit commitment, constrained portfolio construction, and so on \cite{Abbas2024,DeCross2023}. In static unitary circuits, all of the projective readouts are deferred to the end of execution. Conversely, dynamic quantum circuits support mid-circuit ancilla measurements, active qubit resets, and conditional classical feed-forward operations within the physical coherence time $T_{2}$ \cite{DeCross2023,Graham2023}. As a result, the use of these mid-circuit operations to construct non-unitary ground-state attractors for NP-hard constrained binary optimization problems has attracted considerable interest across trapped-ion, neutral-atom, and superconducting platforms \cite{Ebadi2022,Evered2023}.

In the literature, existing gate-model quantum optimization schemes can be classified into three broad categories, namely penalty-flattened unitary schemes, subspace-preserving unitary schemes, and unconstrained measurement-driven cooling schemes. In penalty-flattened schemes, such as the quantum approximate optimization algorithm (QAOA) \cite{Farhi2014}, digitized counterdiabatic quantum optimization (DCQO) \cite{Hegade2022,Kipu2023}, and feedback-based quantum optimization (FALQON) \cite{Magann2022}, the hard combinatorial constraints are embedded into a quadratic unconstrained binary optimization (QUBO) Hamiltonian $H_{\mathrm{QUBO}} = H_{\mathrm{obj}} + \mu H_{\mathrm{pen}}$ via a large scalar penalty multiplier $\mu \gg \|H_{\mathrm{obj}}\|$. However, variational outer-loop optimization requires hundreds if not thousands of iterative circuit executions per problem instance and is subject to barren plateaus, and thus incurs a prohibitively expensive quantum-classical round-trip latency. Deterministic Lyapunov feedback laws such as FALQON eliminate the outer-loop training by setting the mixer angles in proportion to the commutator expectation $i\langle [B, H] \rangle$, where $B$ denotes a single-spin transverse-field mixer. However, $B$ and $H_{\mathrm{obj}}$ are real symmetric in the computational basis, while $i[B, H_{\mathrm{obj}}]$ is antisymmetric. Consequently, the first-order expectation $\langle \psi_{\mathcal{F}} | i[B, H_{\mathrm{obj}}] | \psi_{\mathcal{F}} \rangle$ vanishes identically on any real-valued symmetric feasible superposition $|\psi_{\mathcal{F}}\rangle$, and therefore the deterministic first-order feedback stalls at initialization.

Furthermore, the penalty multiplier limits the resolution of ancilla-based energy filtering. Specifically, when a penalized Hamiltonian $H_{\mathrm{QUBO}}$ is coupled to a phase-estimation or weak-measurement filter, the inflated spectral span $\Delta E_{\mathrm{tot}} = \Delta E_{\mathrm{obj}} + \mu P_{\max}$ must be mapped into the non-aliasing phase interval $[0, \pi]$. The evolution time step is thus compressed as $\Theta(\mu^{-1})$, and the single-cycle rejection probability of the low-lying feasible excited states is reduced quadratically as $\Theta(\mu^{-2})$. In addition, the penalty term introduces dense all-to-all $Z_{i}Z_{j}$ couplings.

Conversely, subspace-preserving unitary schemes restrict the state evolution directly to the feasible subspace $\mathcal{F}$ without the need for scalar penalty multipliers. For example, quantum Zeno dynamics (QZD) \cite{Herman2023} confines the QAOA evolution by means of frequent mid-slice projective syndrome measurements, while quantum constrained Hamiltonian optimization \cite{QCHOP2026} and Grover adaptive search (GAS) \cite{Gilliam2021} employ unitary subspace mixers and reversible binary arithmetic comparators, respectively. However, frequent mid-slice projective measurements of multi-qubit constraints introduce a substantial readout latency and spectator-qubit dephasing. Moreover, reversible ripple-carry adder comparators require approximately $28N$ two-qubit entangling gates per oracle reflection on an $N$-qubit register \cite{Gilliam2021}, and thus exceed the coherent gate budget of near-term and early fault-tolerant processors. In addition, purely unitary subspace evolutions have no absorbing ground-state fixed point, and thus exhibit undamped Grover over-rotation when the ground-state degeneracy or the spectral gap is unknown.

Measurement-driven cooling schemes, including quantum digital cooling \cite{Polla2021} and measurement-based imaginary-time evolution (MITE) \cite{Mao2023}, utilize ancilla measurements to induce non-unitary energy dissipation. However, existing digital cooling schemes do not decouple the hard combinatorial constraints from the cooling filter. Moreover, these schemes provide no noise-resilient closed-loop mechanism for tightening the unknown constrained ground-state reference $E_{\mathrm{ref}}^{(t)}$. As a result, the reference is at risk of overshooting the constrained ground-state energy $E_{0}^{\mathcal{F}}$ (i.e., $E_{\mathrm{ref}}^{(t)} > E_{0}^{\mathcal{F}}$) under realistic mid-circuit readout errors, in which case the dark-state protection of the ground state is destroyed.

Accordingly, this paper proposes an accurate, penalty-free, closed-loop dynamic quantum circuit scheme, designated as Zeno-decoupled gauge-ratchet optimization (Z-GRO), which addresses the limitations described above without outer-loop parameter optimization. In the proposed approach, the hard constraints are decoupled from the objective phase cooling by interleaving a reversible feasibility phase operator $U_{\mathcal{F}}(\varphi)$ within symmetric dynamical-decoupling echoes \cite{Facchi2004}, while the initial stall of symmetric feedback is overcome via a self-calibrated interaction-correlated commutator gauge operator $G_{\mathrm{corr}}$. Directional dissipation toward the constrained ground manifold is then realized by an ancilla spectral ratchet, in which a threshold phase offset $-\tau_{t}E_{\mathrm{ref}}^{(t)}$ is encoded with zero two-qubit gates and conditional mid-circuit feed-forward bypasses the entangling tunneling gates on the retained branches. To the best of the current author's knowledge, Z-GRO represents the first attempt in the literature to integrate reversible phase-echoed gauge tunneling, conditional mid-circuit spectral ratcheting, and sign-invariant rollback threshold control into a unified dynamic-circuit optimization architecture. The performance of Z-GRO is evaluated numerically using exact density-matrix evolution and matrix-free quantum trajectory unraveling for frustrated constrained Sherrington--Kirkpatrick (CSK) spin-glass and quadratic knapsack problem (QKP) ensembles. In the simulations, the performance of Z-GRO is compared to that of five existing quantum schemes, namely penalty-free QZD \cite{Herman2023}, variational QAOA \cite{Farhi2014}, linear-schedule Trotterized quantum annealing (QAOA-TQA), FALQON \cite{Magann2022}, and MITE \cite{Mao2023}, as well as classical random sampling baselines. The results show that Z-GRO achieves a higher optimal ground-state probability and a shorter time-to-solution than all five existing quantum schemes across both ensembles, while achieving the lowest exponential scaling base on the spin-glass ensemble. In addition, it is shown that Z-GRO maintains a higher optimal probability than the noisy existing schemes under two-qubit depolarizing gate errors and is robust toward mid-circuit readout errors. The main contributions of this paper are five-fold:
\begin{enumerate}\setlength{\itemsep}{2pt}
  \item A penalty-free confinement mechanism based on echoed reversible feasibility phase operators $U_{\mathcal{F}}(\varphi)$ is formulated. The mechanism cancels the first-order off-block transitions into infeasible states without mid-slice projective measurements, and thus eliminates the $\Theta(\mu^{-2})$ spectral aliasing inherent in penalized QUBO cooling.
  \item An interaction-correlated commutator gauge operator $G_{\mathrm{corr}}$, calibrated via echoed Ehrenfest curvature readouts, is introduced to guarantee a strictly negative initial energy slope on symmetric feasible superpositions. The operator is fused with correlated transverse mixing into a single echoed generator $K_{\mathrm{tunnel}}$, which halves the finite-slice echo overhead.
  \item An exact completely positive trace-preserving (CPTP) binomial ancilla spectral ratchet channel is established on the basis of a zero-two-qubit-gate threshold phase offset and conditional feed-forward gate bypassing. The channel achieves an analytical $\Theta(\mu^{2})$ discrimination gain over penalized cooling and suppresses the entangling-gate decoherence of the cooled dark states.
  \item A closed-loop sign-invariant threshold controller with automatic overshoot rollback is derived based on survival-filtered Ramsey quadratures, in which the sign of the expectation remains non-negative below the ground-state energy for all symmetric mid-circuit readout error rates below $50\%$. The controller combines automatic overshoot rollback toward the linear-programming bound with circular dispersion-penalized threshold advances and dynamic majority syndrome shielding.
  \item Shallow hardware-native compilation schemes are developed for trapped-ion, neutral-atom, and superconducting processors, including a zero-ancilla single-pulse global M{\o}lmer--S{\o}rensen shell reflection and an epoch-buffered neutral-atom readout pool. In addition, numerical scaling and noise evaluations are performed for $N = 6$ to $14$ qubits, and the entangling gate counts are evaluated for $N = 20$ to $50$ qubits.
\end{enumerate}

The remainder of this paper is organized as follows. Section~\ref{sec:formulation} reviews the related quantum optimization schemes and formulates the constrained binary optimization problem together with the spectral aliasing analysis. Section~\ref{sec:subspace} presents the reversible feasibility phase echo and the interaction-correlated gauge tunneling mechanism. Section~\ref{sec:ratchet} describes the binomial ancilla spectral ratchet and the closed-loop sign-invariant threshold control law. Section~\ref{sec:compilation} describes the shallow platform-native circuit compilation for three hardware platforms. Section~\ref{sec:evaluation} presents and discusses the numerical simulation results. Finally, Sec.~\ref{sec:conclusion} provides some brief concluding remarks. The detailed proofs and implementation parameters are given in Appendices~\ref{app:aliasing}--\ref{app:baselines}.

\section{Related Work and Problem Formulation}
\label{sec:formulation}

\subsection{Related Quantum Optimization Schemes}

Penalty-based variational and counterdiabatic schemes encode constrained combinatorial problems by penalizing the constraint violations within the cost Hamiltonian. Farhi et al. \cite{Farhi2014} proposed QAOA, in which alternating cost and transverse-field mixer unitaries are tuned by a classical outer-loop optimizer. Hegade et al. \cite{Hegade2022,Kipu2023} developed DCQO by incorporating nested-commutator counterdiabatic driving terms into digitized adiabatic schedules. Barkoutsos et al. \cite{Barkoutsos2020} presented a classical end-of-circuit threshold sorting method for improving the variational convergence. Magann et al. \cite{Magann2022} proposed FALQON, in which the mixer angles are assigned deterministically via qubit-measurement feedback without a classical parameter search. However, all of these schemes rely on QUBO penalty flattening when general inequality or conflict constraints are present, and thus suffer from a dense two-qubit coupling overhead and a compressed evolution time step.

Subspace-confinement schemes avoid penalty multipliers by restricting the state evolution to valid bitstrings. Gilliam et al. \cite{Gilliam2021} developed GAS for constrained polynomial binary optimization, in which the objective thresholds and constraints are encoded into coherent reversible arithmetic oracles. Herman et al. \cite{Herman2023} proposed QZD optimization, in which projective ancilla syndrome measurements are interleaved repeatedly across the circuit slices to suppress leakage out of the feasible subspace. However, GAS requires deep ripple-carry comparator trees ($28N$ two-qubit gates per reflection) and exhibits unitary Grover over-rotation. Moreover, QZD incurs a substantial mid-slice readout latency, and the evolution within the feasible subspace remains purely unitary.

Measurement-driven dissipative schemes introduce non-unitary cooling by means of ancilla coupling. Polla et al. \cite{Polla2021} investigated quantum digital cooling, in which the excitations of a many-body system are transferred into a resettable fridge qubit. Mao et al. \cite{Mao2023} proposed MITE based on weak measurements followed by deterministic conditional unitary corrections. Similarly, Shen et al. \cite{Shen2019} presented a binary-tree synthesis of generic Kraus channels using a resettable ancilla. However, none of these schemes decouples the hard combinatorial constraints from the ancilla phase filter. In addition, none of the schemes provides an analytical sign-invariant threshold rollback law under mid-circuit measurement bit-flip noise.

\subsection{System Model and Optimization Problem Formulation}

Consider an $N$-qubit quantum register that encodes the binary decision vector $z = (z_{1}, z_{2}, \dots, z_{N}) \in \{0, 1\}^{N}$ in the computational basis state $|z\rangle$, where the Pauli-$Z$ operator $Z_{i}$ acts on qubit $i$ ($i = 1, 2, \dots, N$) such that $z_{i} = (1 - Z_{i})/2$. The target optimization problem consists of a quadratic binary objective function $E_{\mathrm{obj}}(z)$ subject to $M_{\mathrm{lin}}$ weighted linear inequality constraints and a conflict edge set $E_{\mathrm{conf}}$. It is assumed that the following three conditions hold:
\begin{enumerate}\setlength{\itemsep}{2pt}
  \item The unpenalized objective function is represented by a diagonal Ising Hamiltonian $H_{\mathrm{obj}}$ acting on the $N$-qubit data register, i.e.,
  \begin{equation}
    H_{\mathrm{obj}} = c_{0}I + \sum_{i=1}^{N} h_{i}Z_{i} + \sum_{(i,j)\in E_{\mathrm{obj}}} J_{ij}Z_{i}Z_{j},
    \label{eq:h_obj}
  \end{equation}
  where $c_{0} \in \mathbb{R}$ is a scalar energy offset, $h_{i} \in \mathbb{R}$ is the local longitudinal field on qubit $i$, $J_{ij} \in \mathbb{R}$ is the pairwise interaction weight on edge $(i, j)$ of the objective graph $E_{\mathrm{obj}}$, and $I$ is the $2^{N} \times 2^{N}$ identity operator.
  \item The feasible set $\mathcal{F} \subset \{0, 1\}^{N}$ is non-empty and is defined by the hard constraints as
  \begin{equation}
    \mathcal{F} = \Bigl\{ z \in \{0,1\}^{N} : \sum_{i=1}^{N} w_{m,i}z_{i} \in [L_{m}, U_{m}],\; z_{u}z_{v} = 0 \Bigr\},
    \label{eq:feas_set}
  \end{equation}
  for all $m \in \{1, \dots, M_{\mathrm{lin}}\}$ and all $(u, v) \in E_{\mathrm{conf}}$, where $w_{m,i} \in \mathbb{Z}_{\ge 0}$ is an integer resource weight and $[L_{m}, U_{m}]$ is the admissible capacity interval of constraint $m$. The orthogonal projectors onto $\mathcal{F}$ and its infeasible complement $\mathcal{F}^{\perp}$ are denoted by $\Pi_{\mathcal{F}} = \sum_{z \in \mathcal{F}} |z\rangle\langle z|$ and $\Pi_{\mathcal{F}^{\perp}} = I - \Pi_{\mathcal{F}}$, respectively.
  \item The quantum processor comprises the $N$-qubit data register, an ancilla register of $k_{\mathrm{bus}} \ge 1$ qubits that can be measured and reset in mid-circuit, and real-time digital control circuitry (e.g., an FPGA, an ASIC, or a cryogenic CMOS controller) that is capable of conditional gate execution within the coherence time $T_{2}$.
\end{enumerate}

The constrained optimal binary configuration $z^{*}$ is determined by solving the following combinatorial optimization problem:
\begin{align}
  \mathrm{(P1)} \quad E_{0}^{\mathcal{F}} := \min_{z \in \{0,1\}^{N}} \quad & E_{\mathrm{obj}}(z) = \langle z | H_{\mathrm{obj}} | z \rangle \label{eq:p1_obj}\\
  \text{subject to} \; & \sum_{i=1}^{N} w_{m,i}z_{i} \ge L_{m}, \; \forall m \in \{1,\dots,M_{\mathrm{lin}}\}, \label{eq:p1_c1}\\
  & \sum_{i=1}^{N} w_{m,i}z_{i} \le U_{m}, \; \forall m \in \{1,\dots,M_{\mathrm{lin}}\}, \label{eq:p1_c2}\\
  & z_{u}z_{v} = 0, \quad \forall (u,v) \in E_{\mathrm{conf}}. \label{eq:p1_c3}
\end{align}
The objective of Problem (P1) is to minimize the physical or operational cost $E_{\mathrm{obj}}(z)$ in Eq.~(\ref{eq:p1_obj}). The constraints in Eqs.~(\ref{eq:p1_c1}) and (\ref{eq:p1_c2}) enforce the lower and upper resource capacity bounds, respectively, while the constraint in Eq.~(\ref{eq:p1_c3}) prohibits the simultaneous activation of mutually conflicting decision variables. Problem (P1) belongs to the class of NP-hard constrained binary quadratic programs when the couplings $J_{ij}$ are frustrated.

In conventional penalty-based schemes, Problem (P1) is converted into the unconstrained QUBO problem
\begin{equation}
  \mathrm{(P2)} \quad \min_{z \in \{0,1\}^{N}} \; E_{\mathrm{QUBO}}(z) = E_{\mathrm{obj}}(z) + \mu E_{\mathrm{pen}}(z),
  \label{eq:p2_qubo}
\end{equation}
where $E_{\mathrm{pen}}(z) \ge 0$ is a penalty function that vanishes if and only if $z \in \mathcal{F}$, and $\mu > \Delta E_{\mathrm{obj}}$ is a penalty multiplier that exceeds the feasible objective span $\Delta E_{\mathrm{obj}} := \max_{z\in\mathcal{F}} E_{\mathrm{obj}}(z) - E_{0}^{\mathcal{F}}$. Note that $E_{\mathrm{pen}}(z) \ge 1$ for $z \in \mathcal{F}^{\perp}$, and the maximum penalty is denoted by $P_{\max} := \max_{z} E_{\mathrm{pen}}(z)$. Consider the case in which $H_{\mathrm{QUBO}} = H_{\mathrm{obj}} + \mu H_{\mathrm{pen}}$ is coupled to an interferometric ancilla phase check $\mathrm{ctrl}_{a}\text{-}\exp(-i\tau H)$ with a threshold reference of $E_{\mathrm{ref}} \le E_{0}^{\mathcal{F}}$. To prevent trigonometric wrapping, in which the high-energy infeasible states are aliased back onto the low-energy dark states, the evolution time $\tau$ must satisfy $\tau (\max_{z} E_{\mathrm{QUBO}}(z) - E_{\mathrm{ref}}) \le \pi$. The following proposition then holds.

\begin{proposition}[Penalty-Induced Spectral Aliasing]
\label{prop:aliasing}
Let $\delta E_{1} := E_{1}^{\mathcal{F}} - E_{0}^{\mathcal{F}} > 0$ denote the spectral gap between the first feasible excited state and the constrained ground state of Problem (P1), and let $E_{\mathrm{ref}} = E_{0}^{\mathcal{F}}$. In addition, let $\tau_{\mathrm{pen}} = \pi / (\Delta E_{\mathrm{obj}} + \mu P_{\max})$ and $\tau_{\mathrm{Z\text{-}GRO}} = \pi / \Delta E_{\mathrm{obj}}$ denote the non-aliasing evolution times under the penalized scaling and the penalty-free scaling, respectively. Under the penalized scaling, the single-check ancilla rejection probability $p_{\mathrm{rej}}(E_{1}^{\mathcal{F}}) = \sin^{2}(\tau \delta E_{1} / 2)$ is reduced quadratically with $\mu$, and the discrimination gain of the penalty-free scaling is given by
\begin{equation}
  \mathcal{G}_{\mathrm{kick}}(\mu) := \frac{\sin^{2}(\tau_{\mathrm{Z\text{-}GRO}}\delta E_{1}/2)}{\sin^{2}(\tau_{\mathrm{pen}}\delta E_{1}/2)} = \left(1 + \frac{\mu P_{\max}}{\Delta E_{\mathrm{obj}}}\right)^{2} + \mathcal{O}(\delta E_{1}^{2}).
  \label{eq:kick_gain}
\end{equation}
\end{proposition}

The proof of Proposition~\ref{prop:aliasing} is given in Appendix~\ref{app:aliasing}. As shown in Fig.~\ref{fig:arch}(a), the conversion of the constraints into Problem (P2) compresses the feasible excitations toward zero phase angle, and thus suppresses the dissipative cooling by orders of magnitude. To anchor Z-GRO without Problem (P2) or prior knowledge of $E_{0}^{\mathcal{F}}$, a classical pre-processing step solves a polynomial-time McCormick linear-programming (LP) relaxation, in which linearized edge variables $y_{ij} \in [0, 1]$ are introduced for $(i,j) \in E_{\mathrm{obj}}$, i.e.,
\begin{align}
  \mathrm{(P3)} \quad E_{\mathrm{LP}} := \min_{z, y} \quad & \tilde{c}_{0} + \sum_{i=1}^{N}\tilde{h}_{i}z_{i} + \sum_{(i,j)\in E_{\mathrm{obj}}}\tilde{J}_{ij}y_{ij} \label{eq:p3_lp}\\
  \text{subject to} \; & (\ref{eq:p1_c1}),\; (\ref{eq:p1_c2}),\; z_{u} + z_{v} \le 1\;\forall(u,v)\in E_{\mathrm{conf}}, \nonumber\\
  & 0 \le y_{ij} \le z_{i},\; z_{i}+z_{j}-1 \le y_{ij} \le z_{j}, \nonumber
\end{align}
where $(\tilde{c}_{0}, \tilde{h}_{i}, \tilde{J}_{ij})$ are the binary-form coefficients of Eq.~(\ref{eq:h_obj}). Similarly, maximizing the linearized objective in Eq.~(\ref{eq:p3_lp}) over the identical McCormick constraints yields a companion upper bound, or the closed-form $1$-norm bound $c_{0} + \sum_{i=1}^{N}|h_{i}| + \sum_{(i,j)\in E_{\mathrm{obj}}}|J_{ij}|$ if maximization is omitted. Both linear programs are solved in polynomial time via interior-point methods, and incorporating a small spectral safety margin (Appendix~\ref{app:baselines}) certifies $E_{\mathrm{LP}} \le E_{0}^{\mathcal{F}} \le E_{\mathrm{seed}} \le \max_{z \in \mathcal{F}} E_{\mathrm{obj}}(z) \le E_{\mathrm{ub}}$, where $E_{\mathrm{seed}} = E_{\mathrm{obj}}(z_{\mathrm{seed}})$ is the energy of a greedy feasible seed $z_{\mathrm{seed}} \in \mathcal{F}$.

\begin{figure*}[!tp]
  \centering
  \includegraphics[width=\textwidth]{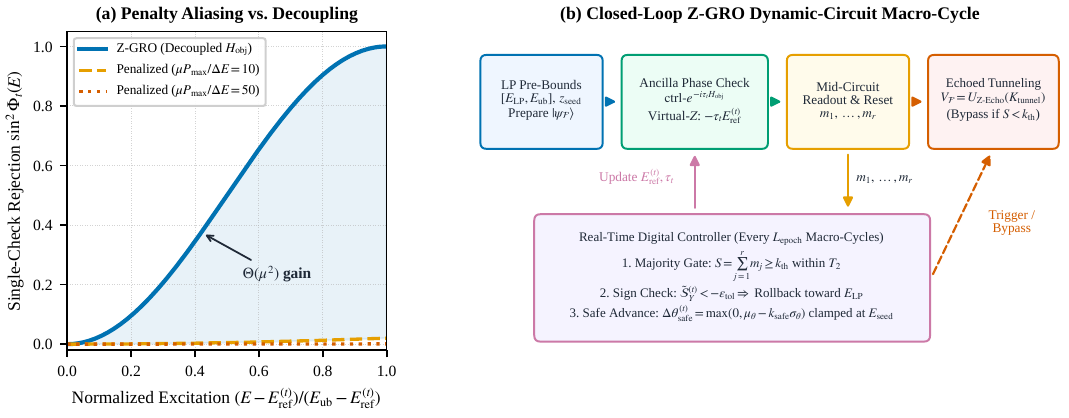}
  \caption{Conceptual comparison and closed-loop dynamic-circuit architecture of Z-GRO. (a) Single-check ancilla rejection probability $\sin^{2}\Phi_{t}(E)$ versus the normalized feasible excitation energy, showing the $\Theta(\mu^{-2})$ spectral compression of the penalized QUBO formulation relative to the penalty-free Z-GRO scaling. (b) Schematic of the closed-loop macro-cycle, comprising the LP pre-bounds, the zero-two-qubit-gate virtual-$Z$ threshold phase encoding, the conditional mid-circuit feed-forward bypassing of $V_{\mathcal{F}}$, and the epoch-level survival-filtered quadrature threshold control.}
  \label{fig:arch}
\end{figure*}

\section{Coherent Subspace Confinement and Echoed Gauge Tunneling}
\label{sec:subspace}

\subsection{Reversible Feasibility Phase Operator and Zeno Echo}

To confine the state evolution to $\mathcal{F}$ without mid-slice projective measurements, Z-GRO utilizes a reversible unitary feasibility phase operator $U_{\mathcal{F}}(\varphi)$, which imparts a non-trivial suppression phase $\varphi \in (0, 2\pi)$ to the infeasible subspace $\mathcal{F}^{\perp}$, i.e.,
\begin{equation}
  U_{\mathcal{F}}(\varphi) := \Pi_{\mathcal{F}} + e^{i\varphi}\Pi_{\mathcal{F}^{\perp}}, \quad R_{\mathcal{F}} := U_{\mathcal{F}}(\pi) = 2\Pi_{\mathcal{F}} - I.
  \label{eq:u_feas}
\end{equation}
Any Hermitian tunneling generator $K$ can be decomposed as $K = K_{\parallel} + K_{\perp}$, where $K_{\parallel} = \Pi_{\mathcal{F}}K\Pi_{\mathcal{F}} + \Pi_{\mathcal{F}^{\perp}}K\Pi_{\mathcal{F}^{\perp}}$ describes the block-diagonal dynamics within the feasible and infeasible subspaces, and $K_{\perp} = \Pi_{\mathcal{F}}K\Pi_{\mathcal{F}^{\perp}} + \Pi_{\mathcal{F}^{\perp}}K\Pi_{\mathcal{F}}$ describes the off-block leakage between the two subspaces. In the canonical involutory reflection case of $\varphi = \pi$, the operator $R_{\mathcal{F}}$ satisfies $R_{\mathcal{F}}^{2} = I$, commutes with $K_{\parallel}$ (i.e., $R_{\mathcal{F}}K_{\parallel}R_{\mathcal{F}} = +K_{\parallel}$), and anticommutes with $K_{\perp}$ (i.e., $R_{\mathcal{F}}K_{\perp}R_{\mathcal{F}} = -K_{\perp}$). Interleaving $R_{\mathcal{F}}$ across $M \ge 1$ symmetric dynamical-decoupling echo pairs \cite{Facchi2004} over a step angle $\theta$ yields the echoed unitary
\begin{equation}
  \begin{aligned}
    U_{\mathrm{Z\text{-}Echo}}(K, \theta, M) &:= \\
    &\left[ R_{\mathcal{F}}\exp\!\left(-i\frac{\theta}{2M}K\right)\! R_{\mathcal{F}}\exp\!\left(-i\frac{\theta}{2M}K\right) \right]^{M}.
  \end{aligned}
  \label{eq:zeno_echo}
\end{equation}
More generally, for an $m$-fold cyclic phase $\varphi = 2\pi q / m$ ($1 \le q < m$), each echo pair in Eq.~(\ref{eq:zeno_echo}) is replaced by $\prod_{\ell=1}^{m} U_{\mathcal{F}}(\varphi)^{\ell}\exp[-i\theta K/(mM)]U_{\mathcal{F}}(\varphi)^{-\ell}$. In this case, the first-order leakage is canceled because $\sum_{\ell=0}^{m-1}e^{i\ell\varphi} = 0$.

\begin{theorem}[Coherent Zeno-Echo Leakage Bound]
\label{thm:zeno}
For any Hermitian generator $K$, step angle $\theta \in \mathbb{R}$, echo slice count $M \ge 1$, and initial feasible state $|\psi\rangle \in \mathrm{Ran}(\Pi_{\mathcal{F}})$, the symmetric echoed unitary in Eq.~(\ref{eq:zeno_echo}) cancels all of the first-order transitions into $\mathcal{F}^{\perp}$ and satisfies
\begin{equation}
  \begin{aligned}
    \bigl\| \Pi_{\mathcal{F}^{\perp}} U_{\mathrm{Z\text{-}Echo}}(K, \theta, M)|\psi\rangle \bigr\|_{2} &\le \frac{\theta^{2}\,\|[K_{\parallel}, K_{\perp}]\|_{2}}{4M} \\
    &\quad + \mathcal{O}\!\left(\frac{\theta^{3}\|K\|_{2}^{3}}{M^{2}}\right),
  \end{aligned}
  \label{eq:leak_bound}
\end{equation}
where $\|\cdot\|_{2}$ denotes the spectral operator norm. Consequently, the infeasible leakage probability $P(\mathcal{F}^{\perp})$ decays asymptotically as $\mathcal{O}(M^{-2})$.
\end{theorem}

The proof of Theorem~\ref{thm:zeno} is given in Appendix~\ref{app:aliasing}. As shown in Fig.~\ref{fig:mech}(a), the use of $M = 4$ echo pairs reduces the infeasible population leakage by a factor of $22.8$ relative to that of an unprotected unitary step, without the need for mid-slice projective measurements.

\subsection{Interaction-Correlated Gauge Launch and Fused Tunneling}

The data register is initialized in the uniform feasible superposition $|\psi_{\mathcal{F}}\rangle = |\mathcal{F}|^{-1/2}\sum_{z \in \mathcal{F}}|z\rangle$. The superposition is obtained by preparing $|+\rangle^{\otimes N}$ and performing $2$ to $4$ initial syndrome post-selections, and the seed state $|z_{\mathrm{seed}}\rangle$ is used as a fallback if required. The state $|\psi_{\mathcal{F}}\rangle$ has strictly real amplitudes, and the single-spin mixer $B_{1} = \sum_{i=1}^{N}X_{i}$ and $H_{\mathrm{obj}}$ are both real symmetric in the computational basis. It follows that $\langle \psi_{\mathcal{F}} | i[B_{1}, H_{\mathrm{obj}}] | \psi_{\mathcal{F}} \rangle = 0$, and thus standard Lyapunov feedback stalls. To induce an immediate coherent descent across the frustrated barriers, and to enable correlated two-spin flips across the parity constraints, Z-GRO constructs an interaction-correlated driver $B_{\mathrm{corr}}$ and the corresponding counterdiabatic commutator gauge operator $G_{\mathrm{corr}}$ as
\begin{equation}
  B_{\mathrm{corr}} = \sum_{i=1}^{N} X_{i} + \kappa \sum_{(u,v)\in \mathcal{P}_{\mathrm{top}\text{-}N}} X_{u}X_{v}, \quad G_{\mathrm{corr}} := i\bigl[B_{\mathrm{corr}},\, H_{\mathrm{obj}}\bigr],
  \label{eq:g_corr}
\end{equation}
where $\kappa > 0$ is a two-body mixing weight and $\mathcal{P}_{\mathrm{top}\text{-}N}$ is the set of up to $N$ objective edges with the largest coupling magnitudes $|J_{uv}|$. Since $iG_{\mathrm{corr}}$ is real antisymmetric, $G_{\mathrm{corr}}$ is a purely imaginary Hermitian operator, and thus breaks the time-reversal symmetry at $t = 0$.

\begin{proposition}[Strict Gauge Energy Descent]
\label{prop:gauge}
For the uniform feasible superposition $|\psi_{\mathcal{F}}\rangle = |\mathcal{F}|^{-1/2}\sum_{z \in \mathcal{F}}|z\rangle$ and any echo slice count $M \ge 1$, suppose that $(B_{\mathrm{corr}})_{z,z'} > 0$ connects at least one pair of feasible states $z, z' \in \mathcal{F}$ with $E_{\mathrm{obj}}(z) \neq E_{\mathrm{obj}}(z')$. Then the evolution under $U_{\mathrm{Z\text{-}Echo}}(G_{\mathrm{corr}}, \eta, M)$ yields a strictly negative initial objective energy slope, i.e.,
\begin{equation}
  \begin{aligned}
    &\frac{d}{d\eta}\Bigl\langle \psi_{\mathcal{F}}\Bigm|U_{\mathrm{Z\text{-}Echo}}^{\dagger}(G_{\mathrm{corr}},\eta,M)H_{\mathrm{obj}} \\
    &\quad \times U_{\mathrm{Z\text{-}Echo}}(G_{\mathrm{corr}},\eta,M)\Bigm|\psi_{\mathcal{F}}\Bigr\rangle\Big|_{\eta=0} \\
    &\quad = -\frac{1}{|\mathcal{F}|}\sum_{z,z'\in\mathcal{F}}(B_{\mathrm{corr}})_{z,z'}\bigl(E_{\mathrm{obj}}(z') - E_{\mathrm{obj}}(z)\bigr)^{2} < 0.
  \end{aligned}
  \label{eq:gauge_slope}
\end{equation}
Moreover, evaluating $\langle H_{\mathrm{obj}} \rangle$ at the three echoed probe angles $\{0, +\delta, -\delta\}$ yields $g_{1} = [E(+\delta) - E(-\delta)]/(2\delta)$ and $g_{2} = [E(+\delta) - 2E(0) + E(-\delta)]/\delta^{2}$, from which the calibrated step is $\eta^{*} = \mathrm{clip}(-g_{1}/g_{2},\allowbreak -\eta_{\max},\allowbreak +\eta_{\max})$ when $g_{2} > \epsilon_{\mathrm{curv}}$ and falls back to the fixed-sign step $\eta^{*} = -\eta_{\mathrm{fb}}\,\mathrm{sgn}(g_{1})$ when $g_{2} \le \epsilon_{\mathrm{curv}}$.
\end{proposition}

\begin{figure*}[!tp]
  \centering
  \includegraphics[width=\textwidth]{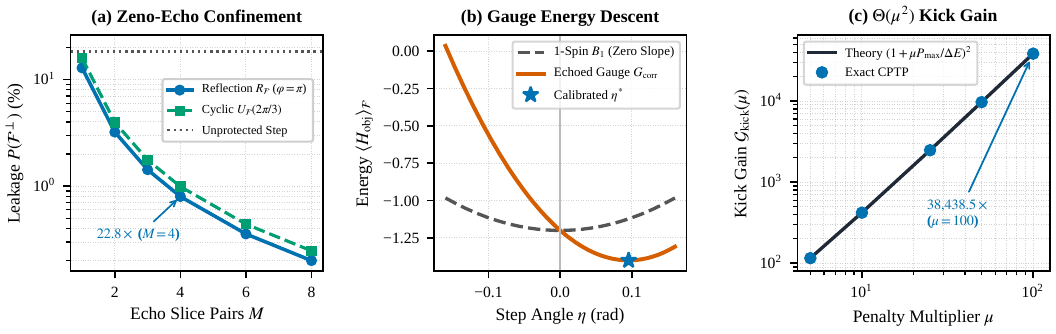}
  \caption{Quantitative verification of the three underlying physical mechanisms of Z-GRO. (a) Infeasible subspace leakage $P(\mathcal{F}^{\perp})$ versus the number of echo slice pairs $M$, showing the $\mathcal{O}(M^{-2})$ suppression ($22.8\times$ reduction at $M = 4$). (b) Objective energy profile $\langle H_{\mathrm{obj}} \rangle_{\mathcal{F}}$ versus the step angle $\eta$ on $|\psi_{\mathcal{F}}\rangle$, showing the zero initial slope under $B_{1}$ and the strictly negative descent under $G_{\mathrm{corr}}$ with the closed-form minimum $\eta^{*}$. (c) Exact CPTP simulation results and the analytical formula in Eq.~(\ref{eq:kick_gain}) for the single-check kick gain $\mathcal{G}_{\mathrm{kick}}(\mu)$ up to $\mu = 100$ ($38{,}438.5\times$).}
  \label{fig:mech}
\end{figure*}

The proof of Proposition~\ref{prop:gauge} is given in Appendix~\ref{app:gauge}. The energy descent under $G_{\mathrm{corr}}$ and the calibrated step $\eta^{*}$ are shown in Fig.~\ref{fig:mech}(b). Following $n_{\mathrm{launch}} \in \{1, 2, 3\}$ coherent gauge launch steps, the subsequent dissipative ratchet cycles employ a single fused Hermitian tunneling generator, i.e.,
\begin{equation}
  K_{\mathrm{tunnel}} = \beta_{0}B_{\mathrm{corr}} + \eta_{\mathrm{rej}}G_{\mathrm{corr}}, \quad V_{\mathcal{F}} := U_{\mathrm{Z\text{-}Echo}}(K_{\mathrm{tunnel}}, 1, M),
  \label{eq:fused_vf}
\end{equation}
where $\beta_{0}$ and $\eta_{\mathrm{rej}}$ are the fixed mixing and gauge amplitudes, respectively. The fusion of $B_{\mathrm{corr}}$ and $G_{\mathrm{corr}}$ in Eq.~(\ref{eq:fused_vf}) halves the number of $R_{\mathcal{F}}$ reflections per rejection event compared to separate mixer and gauge loops.

\section{Dissipative Ancilla Spectral Ratchet and Closed-Loop Control}
\label{sec:ratchet}

\subsection{Zero-Two-Qubit-Gate Threshold Offset and Binomial Ratchet}

To convert the coherent feasible exploration into directional ground-state convergence, each macro-cycle $t$ ($t = 1, 2, \dots, T$) couples the data register to a reusable ancilla. The ancilla is initialized in the state $|+\rangle_{a} = H|0\rangle_{a}$ and is coupled to the data register by the controlled objective evolution $\mathrm{ctrl}_{a}\text{-}\exp(-i\tau_{t}H_{\mathrm{obj}})$, where $\tau_{t} = \pi / (E_{\mathrm{ub}} - E_{\mathrm{ref}}^{(t)})$ is normalized by the certified upper bound $E_{\mathrm{ub}}$ in Sec.~\ref{sec:formulation}. An ancilla phase rotation $R_{z}(-\tau_{t}E_{\mathrm{ref}}^{(t)})$ is then applied, followed by Hadamard interferometry and a $Z$-basis projective readout $m_{j} \in \{0, 1\}$. Note that encoding or updating the dynamic threshold $-\tau_{t}E_{\mathrm{ref}}^{(t)}$ requires zero two-qubit entangling gates. Specifically, the threshold can be realized equivalently by means of four physical implementations, namely (i) a single-qubit $R_{z}(-\tau_{t}E_{\mathrm{ref}}^{(t)})$ pulse on the ancilla, (ii) a zero-duration software-tracked virtual-$Z$ local-oscillator frame shift \cite{McKay2017}, (iii) a rotated ancilla measurement basis $\cos(\tau_{t}E_{\mathrm{ref}}^{(t)})X_{a} + \sin(\tau_{t}E_{\mathrm{ref}}^{(t)})Y_{a}$, or (iv) a shift of the scalar identity term $c_{0} \leftarrow c_{0} - E_{\mathrm{ref}}^{(t)}$ in Eq.~(\ref{eq:h_obj}). Under any of these implementations, a single ancilla check induces the diagonal Kraus operators $M_{0} = \cos\Phi_{t}$ ($m_{j} = 0$, retention) and $M_{1} = \sin\Phi_{t}$ ($m_{j} = 1$, excitation), where
\begin{equation}
  \begin{aligned}
    \Phi_{t}(z) &= \frac{\tau_{t}}{2}\Bigl(E_{\mathrm{obj}}(z) - E_{\mathrm{ref}}^{(t)}\Bigr) \in \left[0, \frac{\pi}{2}\right], \\
    &\quad \forall z \in \mathcal{F} \text{ when } E_{\mathrm{ref}}^{(t)} \le E_{0}^{\mathcal{F}}.
  \end{aligned}
  \label{eq:phi_t}
\end{equation}

The ancilla phase check and active reset are repeated $r \ge 1$ times to yield the syndrome bits $(m_{1}, \dots, m_{r})$ with an integer sum of $S = \sum_{j=1}^{r} m_{j}$. The real-time control circuitry then evaluates the binomial rejection predicate $S \ge k_{\mathrm{th}}$ for $k_{\mathrm{th}} \in \{1, \dots, r\}$. If $S \ge k_{\mathrm{th}}$, the echoed tunneling unitary $V_{\mathcal{F}}$ in Eq.~(\ref{eq:fused_vf}) is triggered within $T_{2}$. Otherwise (i.e., $S < k_{\mathrm{th}}$), $V_{\mathcal{F}}$ is bypassed completely, and thus the cooled states are spared the decoherence of the two-qubit entangling gates. Tracing out the reset ancilla yields the exact CPTP binomial spectral ratchet channel
\begin{equation}
  \mathcal{E}_{(r, k_{\mathrm{th}})}(\rho) = \sum_{s=0}^{k_{\mathrm{th}}-1}\binom{r}{s}K_{r,s}\rho K_{r,s}^{\dagger} + V_{\mathcal{F}}\!\left(\sum_{s=k_{\mathrm{th}}}^{r}\binom{r}{s}K_{r,s}\rho K_{r,s}^{\dagger}\right)\!V_{\mathcal{F}}^{\dagger},
  \label{eq:binomial_cptp}
\end{equation}
where $K_{r,s} := \cos^{r-s}(\Phi_{t})\sin^{s}(\Phi_{t})$.

\begin{theorem}[Dark-State Stationarity and Contraction]
\label{thm:ratchet}
Assume ideal feasible confinement (i.e., $\Pi_{\mathcal{F}^{\perp}}V_{\mathcal{F}}\Pi_{\mathcal{F}} = 0$) and ergodic mixing of $V_{\mathcal{F}}$ over $\mathcal{F}$ (i.e., $\langle z^{*} | V_{\mathcal{F}} | z \rangle \neq 0$). When $E_{\mathrm{ref}}^{(t)} = E_{0}^{\mathcal{F}}$, any density operator $\rho^{*}$ supported exclusively on the optimal feasible ground manifold $\mathcal{F}^{*} = \{z \in \mathcal{F} : E_{\mathrm{obj}}(z) = E_{0}^{\mathcal{F}}\}$ satisfies $\Phi_{t}(z^{*}) = 0$ and $\mathcal{E}_{(r, k_{\mathrm{th}})}(\rho^{*}) = \rho^{*}$ for all $1 \le k_{\mathrm{th}} \le r$. In other words, $\mathcal{F}^{*}$ forms an absorbing stationary manifold, and the asymptotic per-cycle contraction factor of the excited states is given by $1 - \gamma_{\min}\sum_{s=k_{\mathrm{th}}}^{r}\binom{r}{s}\sin^{2s}(\Phi_{\min})\cos^{2(r-s)}(\Phi_{\min}) < 1$, where $\Phi_{\min} = \tau_{t}\delta E_{1}/2$ and $\gamma_{\min} > 0$.
\end{theorem}

The proof of Theorem~\ref{thm:ratchet} is given in Appendix~\ref{app:ratchet}. Theorem~\ref{thm:ratchet} shows that a threshold overshoot (i.e., $E_{\mathrm{ref}}^{(t)} > E_{0}^{\mathcal{F}}$) is destructive. Specifically, if $E_{\mathrm{ref}}^{(t)} > E_{0}^{\mathcal{F}}$, then $\Phi_{t}(z^{*}) < 0$ and $\sin^{2}\Phi_{t}(z^{*}) > 0$. As a result, the true ground state is destabilized and a spurious dark state is created near $E_{\mathrm{ref}}^{(t)}$.

\subsection{Sign-Invariant Threshold Control Law with Overshoot Rollback}

The objective of the threshold controller is to tighten $E_{\mathrm{ref}}^{(t)}$ upward from $E_{\mathrm{LP}}$ toward $E_{0}^{\mathcal{F}}$ without locking above $E_{0}^{\mathcal{F}}$ under a symmetric mid-circuit ancilla readout bit-flip error rate of $\epsilon_{\mathrm{MCM}} \in [0, 0.5)$. To achieve this objective, the digital controller updates $E_{\mathrm{ref}}^{(t)}$ every $L_{\mathrm{epoch}} \ge 1$ macro-cycles using only the trajectories that pass $K_{\mathrm{filt}} \ge 1$ consecutive zero-syndrome checks. The surviving trajectories have dark-filtered weights $w_{\mathrm{dark}}(z) \propto \rho_{t}(z,z)\cos^{2K_{\mathrm{filt}}}\Phi_{t}(z)$, which are concentrated near $E_{0}^{\mathcal{F}}$. For the surviving shots, two Ramsey ancilla quadratures are estimated at the evolution time $\tau_{t}$, namely the cosine quadrature $\tilde{\mathcal{C}}_{X}^{(t)} = \widehat{\mathbb{E}}[1 - 2m_{X}]$ and the sine quadrature $\tilde{\mathcal{S}}_{Y}^{(t)} = \widehat{\mathbb{E}}[1 - 2m_{Y}]$. The sine quadrature is obtained by replacing the final ancilla Hadamard gate with $R_{x}(+\pi/2)$.

\begin{theorem}[Readout-Noise Sign Invariance and Safe Advance]
\label{thm:control}
Under any symmetric mid-circuit ancilla readout bit-flip channel that flips $0 \leftrightarrow 1$ with a probability of $\epsilon_{\mathrm{MCM}} \in [0, 0.5)$, the noisy quadrature expectations satisfy
\begin{equation}
  \mathbb{E}\bigl[\tilde{\mathcal{S}}_{Y}^{(t)}\bigr] = (1 - 2\epsilon_{\mathrm{MCM}})\sum_{z \in \mathcal{F}} w_{\mathrm{dark}}(z)\sin\!\Bigl(\tau_{t}\bigl(E_{\mathrm{obj}}(z) - E_{\mathrm{ref}}^{(t)}\bigr)\Bigr),
  \label{eq:sign_invariance}
\end{equation}
and $\mathbb{E}[\tilde{\mathcal{C}}_{X}^{(t)}] = (1 - 2\epsilon_{\mathrm{MCM}})\sum_{z \in \mathcal{F}} w_{\mathrm{dark}}(z)\cos(\tau_{t}(E_{\mathrm{obj}}(z) - E_{\mathrm{ref}}^{(t)}))$. Since $1 - 2\epsilon_{\mathrm{MCM}} > 0$ and $\tau_{t}(E_{\mathrm{obj}}(z) - E_{\mathrm{ref}}^{(t)}) \in [0, \pi]$ for all $z \in \mathcal{F}$ when $E_{\mathrm{ref}}^{(t)} \le E_{0}^{\mathcal{F}}$, it holds that $\mathbb{E}[\tilde{\mathcal{S}}_{Y}^{(t)}] \ge 0$ for every threshold that has not overshot (i.e., $E_{\mathrm{ref}}^{(t)} \le E_{0}^{\mathcal{F}}$), irrespective of the value of $\epsilon_{\mathrm{MCM}} < 0.5$.
\end{theorem}

The proof of Theorem~\ref{thm:control} is given in Appendix~\ref{app:control}. Based on Theorem~\ref{thm:control}, $\mathbb{E}[\tilde{\mathcal{S}}_{Y}^{(t)}] \ge 0$ holds whenever $E_{\mathrm{ref}}^{(t)} \le E_{0}^{\mathcal{F}}$, which prevents false rollbacks below the ground-state energy. Consequently, observing $\tilde{\mathcal{S}}_{Y}^{(t)} < -\epsilon_{\mathrm{tol}}$ for $\epsilon_{\mathrm{tol}} \ge 0$ provides a sufficient certificate of a threshold overshoot ($E_{\mathrm{ref}}^{(t)} > E_{0}^{\mathcal{F}}$), whereas an overshoot can remain undetected with $\tilde{\mathcal{S}}_{Y}^{(t)} \ge 0$ if ground-state weight in $w_{\mathrm{dark}}(z)$ is small relative to excited feasible states. Whenever $\tilde{\mathcal{S}}_{Y}^{(t)} < -\epsilon_{\mathrm{tol}}$, the controller performs a convex rollback toward $E_{\mathrm{LP}}$, i.e.,
\begin{equation}
  E_{\mathrm{ref}}^{(t+1)} = (1 - \gamma_{\mathrm{rb}})E_{\mathrm{LP}} + \gamma_{\mathrm{rb}}E_{\mathrm{ref}}^{(t)}, \quad \text{if } \tilde{\mathcal{S}}_{Y}^{(t)} < -\epsilon_{\mathrm{tol}},
  \label{eq:rollback}
\end{equation}
where $\gamma_{\mathrm{rb}} \in (0, 1)$ is the rollback coefficient (e.g., $\gamma_{\mathrm{rb}} = 0.5$). Conversely, when $\tilde{\mathcal{S}}_{Y}^{(t)} \ge -\epsilon_{\mathrm{tol}}$, the controller computes the circular mean phase $\mu_{\theta} = \mathrm{atan2}(\tilde{\mathcal{S}}_{Y}^{(t)}, \tilde{\mathcal{C}}_{X}^{(t)})$, the visibility-compensated coherence length $R_{\mathrm{coh}} = \min(1,\allowbreak \sqrt{(\tilde{\mathcal{C}}_{X}^{(t)})^{2} + (\tilde{\mathcal{S}}_{Y}^{(t)})^{2}} / (1 - 2\epsilon_{\mathrm{MCM}}))$, and the circular dispersion $\sigma_{\theta} = \sqrt{-2\ln R_{\mathrm{coh}}}$. The threshold $E_{\mathrm{ref}}^{(t)}$ is then advanced conservatively by means of a lower-tail safety margin, which is clamped at $E_{\mathrm{seed}}$, i.e.,
\begin{align}
  \Delta\theta_{\mathrm{safe}}^{(t)} &= \max\bigl(0,\; \mu_{\theta} - k_{\mathrm{safe}}\sigma_{\theta}\bigr), \label{eq:delta_theta}\\
  E_{\mathrm{ref}}^{(t+1)} &= (1 - \alpha)E_{\mathrm{ref}}^{(t)} + \alpha \min\!\left(E_{\mathrm{ref}}^{(t)} + \frac{\Delta\theta_{\mathrm{safe}}^{(t)}}{\tau_{t}},\; E_{\mathrm{seed}}\right), \label{eq:cantelli_update}
\end{align}
where $k_{\mathrm{safe}} > 0$ and $\alpha \in (0, 1]$ are the dispersion safety multiplier and the damping gain, respectively.

Under a readout error rate of $\epsilon_{\mathrm{MCM}} > 0$, an ideal ground dark state (i.e., $\Phi_{t}(z^{*}) = 0$) experiences spurious ancilla clicks with a probability of $\epsilon_{\mathrm{MCM}}$ per check. For the setting $(r, k_{\mathrm{th}}) = (3, 1)$, the false-kick probability is $1 - (1 - \epsilon_{\mathrm{MCM}})^{3} = 3\epsilon_{\mathrm{MCM}} + \mathcal{O}(\epsilon_{\mathrm{MCM}}^{2})$. Conversely, the majority setting $(r, k_{\mathrm{th}}) = (3, 2)$ suppresses the false kicks of the ground state quadratically to
\begin{equation}
  P_{\mathrm{false}}^{(3,2)}(\epsilon_{\mathrm{MCM}}) = 3\epsilon_{\mathrm{MCM}}^{2}(1 - \epsilon_{\mathrm{MCM}}) + \epsilon_{\mathrm{MCM}}^{3}.
  \label{eq:majority_shield}
\end{equation}
Accordingly, the controller monitors the dark-branch rejection rate $q_{\mathrm{dark}}^{(t)}$. The $(3, 2)$ setting is engaged when $q_{\mathrm{dark}}^{(t)} > q_{\mathrm{hi}}$ or $\epsilon_{\mathrm{MCM}}$ exceeds a calibrated threshold, the intermediate $(2, 1)$ setting is applied for $q_{\mathrm{lo}} < q_{\mathrm{dark}}^{(t)} \le q_{\mathrm{hi}}$, and the high-gain $(3, 1)$ setting is selected when $q_{\mathrm{dark}}^{(t)} \le q_{\mathrm{lo}}$. In addition, a single projective feasibility syndrome check is performed every $L_{\mathrm{epoch}}$ cycles, which resets any leakage into $\mathcal{F}^{\perp}$ onto $V_{\mathcal{F}}|z_{\mathrm{seed}}\rangle$.

\begin{figure*}[!tp]
  \centering
  \includegraphics[width=\textwidth]{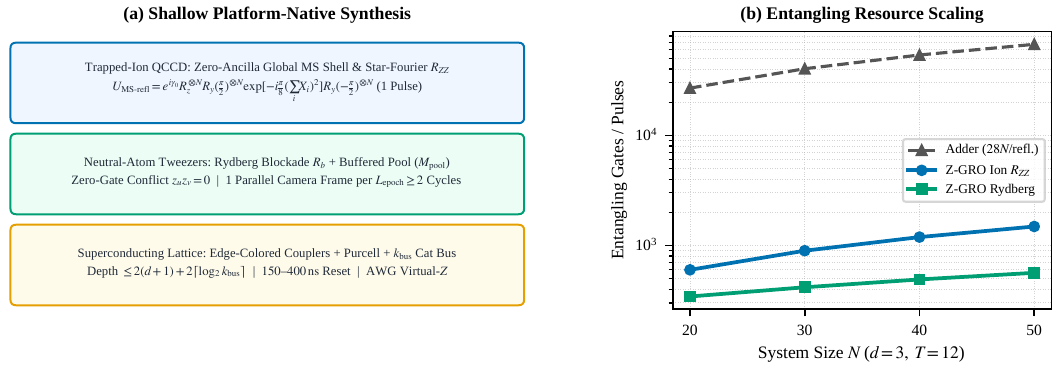}
  \caption{Shallow hardware-native compilation and entangling gate scaling of Z-GRO. (a) Platform-native primitives for trapped-ion QCCD processors (zero-ancilla global MS shell reflection and star-Fourier $R_{ZZ}$ gates), neutral-atom optical tweezers (Rydberg blockade conflict enforcement and epoch-buffered readout pool of $M_{\mathrm{pool}}$ atoms), and superconducting coupler lattices (Purcell-filtered fast reset and GHZ cat-state bus of $k_{\mathrm{bus}}$ qubits). (b) Total number of entangling gates or parallel Rydberg pulses versus the system size $N \in [20, 50]$ ($d = 3$, $T = 12$), compared to that of reversible $28N$-gate adder comparators.}
  \label{fig:hw}
\end{figure*}

\begin{table*}[!tp]
  \centering
  \vspace{-4pt}\caption{Structural comparison of representative constrained and dissipative quantum optimization schemes.}
  \label{tab:taxonomy}
  \footnotesize
  \begin{tabularx}{\textwidth}{@{}lXXXXXXX@{}}
    \toprule
    \textbf{Scheme} & \textbf{Penalty-Free} $\boldsymbol{H_{\mathrm{obj}}}$ & \textbf{Outer-Loop-Free} & \textbf{Ground Attractor} & \textbf{Constraint Gate Cost} & \textbf{Dark-Branch Gate Bypass} & \textbf{Rollback Control} & \textbf{Readout Error Shielding} \\
    \midrule
    QAOA \cite{Farhi2014} & No ($\mu H_{\mathrm{pen}}$) & No (Variational) & No (Unitary) & $\mathcal{O}(N^{2})$ $ZZ$ & No & N/A & N/A \\
    DCQO \cite{Hegade2022,Kipu2023} & No ($\mu H_{\mathrm{pen}}$) & Optional & No (Unitary) & $\mathcal{O}(N^{2})$ $ZZ$ & No & N/A & N/A \\
    FALQON \cite{Magann2022} & No ($\mu H_{\mathrm{pen}}$) & Yes (1st-order) & Stalls on $|\psi_{\mathcal{F}}\rangle$ & $\mathcal{O}(N^{2})$ $ZZ$ & No & N/A & No \\
    QZD \cite{Herman2023} & Yes (Projective) & No (Variational) & No (Unitary) & Frequent MCM & No & N/A & No \\
    GAS \cite{Gilliam2021} & Yes (Oracle) & Yes (Grover) & No (Over-rotates) & $\approx 28N$ 2Q/refl. & No & Classical min & N/A \\
    Digital Cooling \cite{Polla2021} & Unconstrained & Yes & Static bath & N/A & No & None & No \\
    MITE \cite{Mao2023} & No ($\mu H_{\mathrm{pen}}$) & Yes & Heuristic target & $\mathcal{O}(N^{2})$ $ZZ$ & No & None & No \\
    \textbf{Z-GRO (Proposed)} & \textbf{Yes} ($\boldsymbol{U_{\mathcal{F}}(\varphi)}$) & \textbf{Yes} (Closed-loop) & \textbf{Yes} (CPTP fixed pt.) & \textbf{1 MS / Blockade} & \textbf{Yes} (Within $T_{2}$) & \textbf{Yes} ($\boldsymbol{\tilde{\mathcal{S}}_{Y}^{(t)}}$ sign) & \textbf{Yes} ($\boldsymbol{\mathcal{O}(\epsilon_{\mathrm{MCM}}^{2})}$) \\
    \bottomrule
  \end{tabularx}
\end{table*}

\section{Shallow Platform-Native Circuit Compilation}
\label{sec:compilation}

Replacing generic $28N$-gate binary ripple-carry comparators with native multi-body interactions reduces the circuit depth by more than one order of magnitude on the leading quantum processors. The platform-native primitives are summarized in Fig.~\ref{fig:hw}(a), and the structural properties of Z-GRO are compared to those of the existing schemes in Table~\ref{tab:taxonomy}.

\subsection{Trapped-Ion Global MS Shell Reflection and Star-Fourier Gates}

On trapped-ion processors that support global bichromatic M{\o}lmer--S{\o}rensen (MS) interactions \cite{Sorensen1999} and isolated-zone mid-circuit readout in a quantum charge-coupled device (QCCD) architecture \cite{DeCross2023}, band-cardinality constraints of width two take the form $W(z) = \sum_{i=1}^{N}z_{i} \in \{C_{0}, C_{0}+1\}$. Let $\Delta = W(z) - C_{0}$ denote the integer deviation. The binomial integer $\Delta(\Delta - 1)/2$ is even for $\Delta \in \{0, 1\}$ and odd for the neighboring shells $\Delta \in \{-2, -1, +2, +3\}$.

\begin{proposition}[Single-Pulse Global MS Reflection]
\label{prop:ms}
For $\Delta = \sum_{i=1}^{N}z_{i} - C_{0}$, the parity phase $\exp[-i\frac{\pi}{2}\Delta(\Delta - 1)]$ is equal to $+1$ on $\{C_{0}, C_{0}+1\}$ and $-1$ on $\{\Delta \equiv 2, 3 \pmod 4\}$. Moreover, the parity phase is synthesized with zero ancilla qubits using a single global bichromatic MS pulse of angle $\chi = \pi/8$ according to
\begin{equation}
  \begin{aligned}
    U_{\mathrm{MS\text{-}refl}} &= e^{i\gamma_{0}}\!\left(\bigotimes_{i=1}^{N} R_{z}(\vartheta_{0})\right)\! R_{y}\!\left(\frac{\pi}{2}\right)^{\otimes N} \\
    &\quad \times \exp\!\left(-i\frac{\pi}{8}\Bigl(\sum_{i=1}^{N}X_{i}\Bigr)^{2}\right)\! R_{y}\!\left(-\frac{\pi}{2}\right)^{\otimes N},
  \end{aligned}
  \label{eq:ms_refl}
\end{equation}
where $\vartheta_{0} = -\pi(N - 2C_{0} - 1)/2$ and $\gamma_{0}$ denotes a global phase.
\end{proposition}

The proof of Proposition~\ref{prop:ms} is given in Appendix~\ref{app:hardware}. For general weighted linear inequalities of the form $\sum_{i=1}^{N}w_{i}z_{i} \in [L, U]$, the operator $U_{\mathcal{F}}(\varphi)$ is compiled without ripple-carry adders by means of $H_{\mathrm{harm}}$ star-coupled Fourier harmonics $\prod_{i=1}^{N}R_{ZZ}^{(a,i)}(-\alpha_{h}w_{i}/2)$ between a single ancilla $a$ and the data qubits $i$, where $\alpha_{h}$ are real Fourier coefficients.

\subsection{Neutral-Atom Rydberg Blockade and Epoch-Buffered Pool}

On optical-tweezer neutral-atom processors \cite{Ebadi2022,Evered2023}, each pair of conflicting atoms $(u, v) \in E_{\mathrm{conf}}$ is placed within the Rydberg blockade radius $R_{b}$. The blockade natively and continuously prohibits the doubly excited states (i.e., $z_{u}z_{v} = 0$), and thus the conflict constraints are enforced with zero digital reflection gates. However, single-species fluorescence imaging requires approximately $500\,\mu\mathrm{s}$ to several milliseconds per camera exposure \cite{Graham2023}. To eliminate mid-epoch camera stalls, Z-GRO allocates an epoch-buffered ancilla pool of $M_{\mathrm{pool}} = r L_{\mathrm{epoch}}$ spectator atoms. Over $L_{\mathrm{epoch}} \ge 2$ macro-cycles, the syndrome checks are written sequentially into fresh pool atoms via Rydberg controlled-phase gates and local coherent conditional addressing. All of the $M_{\mathrm{pool}}$ ancilla atoms are then transported once per epoch into a shielded readout zone and are imaged in a single parallel fluorescence frame.

\subsection{Superconducting Coupler Lattice and Cat-State Ancilla Bus}

On planar superconducting transmon or fluxonium lattices, the sequential coupling of a single ancilla to an $N$-qubit graph incurs a SWAP routing overhead. To reduce this overhead, a $k_{\mathrm{bus}}$-qubit Greenberger--Horne--Zeilinger (GHZ) cat state $(|0\rangle^{\otimes k_{\mathrm{bus}}} + |1\rangle^{\otimes k_{\mathrm{bus}}})/\sqrt{2}$ is prepared in $\lceil \log_{2}k_{\mathrm{bus}} \rceil$ CNOT layers. The cat state enables disjoint edge-colored interaction subsets of a graph with a maximum degree of $d$ to accumulate the objective phase in parallel across the $k_{\mathrm{bus}}$ ancilla copies. As a result, the phase accumulation requires at most $2(d + 1) + 2\lceil \log_{2}k_{\mathrm{bus}} \rceil$ two-qubit gate layers prior to the cat uncomputation. Purcell-filtered dispersive resonators perform the mid-circuit ancilla readout and active reset within $150\text{--}400\,\mathrm{ns}$, and thus enable $L_{\mathrm{epoch}} = 1$. In addition, room-temperature radio-frequency system-on-chip controllers or cryogenic arbitrary waveform generators (AWGs) update $-\tau_{t}E_{\mathrm{ref}}^{(t)}$ via zero-duration virtual-$Z$ phase registers. Under a two-qubit depolarizing error rate of $\epsilon_{2Q}$, hardware execution can terminate early at a noise-limited stopping depth $T^{*}(\epsilon_{2Q})$ estimated via a coarse depth-sweep calibration before depolarizing heating overtakes ratchet cooling, whereas Sec.~\ref{sec:evaluation} evaluates both the post-hoc trajectory maximum $T^{*}(\epsilon_{2Q})$ and fixed-depth $T = 24$ execution.

\begin{figure*}[!tp]
  \centering
  \includegraphics[width=\textwidth]{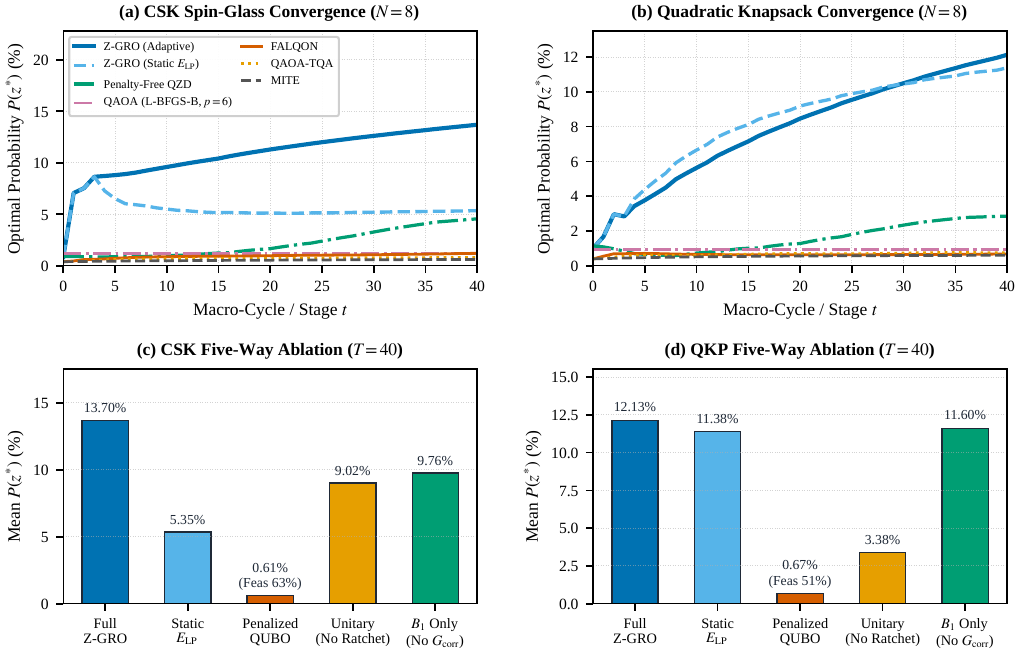}
  \caption{Exact CPTP density-matrix convergence trajectories and five-way ablation results for $N = 8$ ($T = 40$, $M = 3$, $r = 3$, $L_{\mathrm{epoch}} = 4$). (a) Mean optimal probability $P(z^{*})$ over $12$ frustrated CSK spin-glass instances. (b) Mean optimal probability $P(z^{*})$ over $12$ frustrated QKP instances. (c), (d) Five-way ablation results for the CSK and QKP ensembles, respectively, showing the contributions of the adaptive $E_{\mathrm{ref}}^{(t)}$ tracking, the penalty-free $R_{\mathcal{F}}$ echoing, the dissipative ancilla ratcheting, and the correlated commutator gauge $G_{\mathrm{corr}}$.}
  \label{fig:conv_abl}
\end{figure*}

\begin{table*}[!tp]
  \centering
  \caption{Quantitative performance summary of the $N = 8$ exact CPTP benchmarks, the five-way ablations, and the $N = 6\text{--}14$ trajectory scaling ($5$ instances per $N$ per ensemble). Uncertainties ($\pm$) indicate one standard error of the mean ($\sigma/\sqrt{12}$, using population standard deviation $\sigma$) across $12$ instances for $P(z^{*})$ and one ordinary least-squares regression standard error across $N \in \{6, 8, 10, 12, 14\}$ for $b$.}
  \label{tab:results}
  \footnotesize
  \begin{tabularx}{\textwidth}{@{}Xccccccc@{}}
    \toprule
    \textbf{Scheme / Ablation Variant} & \textbf{CSK} $\boldsymbol{P(z^{*})}$ & \textbf{CSK Feas.} & \textbf{QKP} $\boldsymbol{P(z^{*})}$ & \textbf{QKP Feas.} & \textbf{CSK Base} $\boldsymbol{b}$ & \textbf{QKP Base} $\boldsymbol{b}$ & \textbf{Overshoots (}$\boldsymbol{N=8}$\textbf{)} \\
    \midrule
    \textbf{Z-GRO (Adaptive Full)} & \textbf{13.70\% $\pm$ 1.59\%} & \textbf{99.99\%} & \textbf{12.13\% $\pm$ 0.75\%} & \textbf{99.93\%} & \textbf{1.529 $\pm$ 0.050} & \textbf{2.022 $\pm$ 0.064} & \textbf{0 / 24} \\
    Ablation 1: Static LP Bound ($E_{\mathrm{ref}}^{(t)} \equiv E_{\mathrm{LP}}$, $(3, 1)$) & 5.35\% $\pm$ 2.09\% & 99.97\% & 11.38\% $\pm$ 1.84\% & 99.79\% & N/A & N/A & 0 / 24 \\
    Ablation 2: Penalized $H_{\mathrm{QUBO}}$ (Without $R_{\mathcal{F}}$ Echo) & 0.61\% $\pm$ 0.12\% & 63.38\% & 0.67\% $\pm$ 0.05\% & 50.88\% & N/A & N/A & 0 / 24 \\
    Ablation 3: Unitary Echoed Gauge (Without Ratchet) & 9.02\% $\pm$ 1.58\% & 100.00\% & 3.38\% $\pm$ 0.26\% & 100.00\% & N/A & N/A & N/A \\
    Ablation 4: Single-Spin $B_{1}$ Only ($\kappa = 0$, Without $G_{\mathrm{corr}}$) & 9.76\% $\pm$ 1.35\% & 99.99\% & 11.60\% $\pm$ 0.77\% & 99.98\% & N/A & N/A & 0 / 24 \\
    Baseline: Penalty-Free QZD (TQA Schedule) \cite{Herman2023} & 4.55\% $\pm$ 0.73\% & 52.06\% & 2.85\% $\pm$ 0.38\% & 34.60\% & 1.684 $\pm$ 0.039 & 2.003 $\pm$ 0.038 & N/A \\
    Baseline: Variational QAOA (L-BFGS-B, $p = 6$) & 1.20\% $\pm$ 0.13\% & N/A & 0.93\% $\pm$ 0.03\% & N/A & 1.909 $\pm$ 0.116 & 2.076 $\pm$ 0.028 & N/A \\
    Baseline: FALQON ($\mu$-Penalized Lyapunov Feedback) & 1.20\% $\pm$ 0.17\% & N/A & 0.67\% $\pm$ 0.04\% & N/A & 1.980 $\pm$ 0.114 & 2.083 $\pm$ 0.031 & N/A \\
    Baseline: QAOA-TQA (Linear Annealing Schedule) & 0.80\% $\pm$ 0.04\% & N/A & 0.75\% $\pm$ 0.02\% & N/A & 1.999 $\pm$ 0.051 & 2.085 $\pm$ 0.023 & N/A \\
    Baseline: MITE (Penalized Weak-Measurement ITE) & 0.61\% $\pm$ 0.01\% & N/A & 0.60\% $\pm$ 0.01\% & N/A & 2.047 $\pm$ 0.009 & 2.057 $\pm$ 0.012 & N/A \\
    \bottomrule
  \end{tabularx}
\end{table*}

\section{Performance Evaluation}
\label{sec:evaluation}

The performance of the proposed Z-GRO scheme was evaluated by means of a series of numerical simulations based on exact CPTP density-matrix evolution ($\dim = 2^{N}$) and matrix-free fast Walsh--Hadamard quantum trajectory unraveling for up to $N = 14$ qubits ($\dim = 16{,}384$).

\subsection{Simulation Settings and Benchmark Ensembles}

Two NP-hard constrained benchmark classes with frustrated couplings were generated. In the CSK spin-glass ensemble, Gaussian couplings $J_{ij} \sim \mathcal{N}(0, 1/N)$ ($i < j$) and fields $h_{i} \sim \mathcal{N}(0, 0.45^{2})$ were paired with the cardinality interval $\sum_{i=1}^{N}z_{i} \in [\lfloor 0.25N \rfloor, \lfloor 0.50N \rfloor]$ and two disjoint conflict exclusions $z_{1}z_{2} = 0$ and $z_{3}z_{4} = 0$ ($|\mathcal{F}| = 6{,}808$ at $N = 14$). In the QKP ensemble, integer weights $w_{i} \in \{3, \dots, 9\}$, linear values $p_{i} = 0.35w_{i} + \mathcal{N}(0, 0.8^{2})$, and frustrated pairwise synergies $P_{ij} \sim \mathcal{N}(0.15, 1.15^{2}/N)$ were paired with the capacity window $[\lfloor 0.22\sum_{i}w_{i} \rfloor, \lfloor 0.45\sum_{i}w_{i} \rfloor]$ (mean $|\mathcal{F}| = 5{,}747$ at $N = 14$). Each $N = 8$ benchmark comprised $12$ instances per ensemble, while the $N \in \{6, 8, 10, 12, 14\}$ scaling study evaluated $5$ instances per $N$ per ensemble ($25$ CSK and $25$ QKP instances) using $80$ quantum trajectories per instance for $N \le 12$ and $48$ trajectories for $N = 14$ for both Z-GRO and MITE. In the main CSK and QKP runs, $R_{\mathcal{F}} = 2\Pi_{\mathcal{F}} - I$ was simulated as an ideal diagonal phase reflection. To validate Proposition~\ref{prop:ms} directly, an auxiliary $12$-instance constrained spin-glass test at $N = 8$ ($T = 40$, $M = 3$) was performed on the width-two shell $\sum_{i=1}^{8}z_{i} \in \{3, 4\}$ ($C_{0} = 3$). Replacing ideal $R_{\mathcal{F}}$ with the physical single-pulse global MS operator $U_{\mathrm{MS\text{-}refl}}$ in Eq.~(\ref{eq:ms_refl}) achieved $P(z^{*}) = 13.12\% \pm 1.87\%$ ($99.98\%$ feasibility), matching ideal $R_{\mathcal{F}}$ ($13.04\% \pm 1.84\%$).

All outer-loop-free schemes were evaluated over an identical budget of coherent stages, namely $T = 40$ at $N = 8$ ($P_{\mathrm{QZD}} = 10$ outer layers of $m_{\mathrm{sub}} = 4$ projected mixer slices for QZD) and $T = 24$ for $N = 6\text{--}14$ ($P_{\mathrm{QZD}} = 8$, $m_{\mathrm{sub}} = 3$ for QZD). All four penalized baselines encoded violations directly as diagonal hinge-squared penalties $E_{\mathrm{pen}}(z)$ without slack qubits at $\mu = 2\max(E_{\mathrm{ub}} - E_{\mathrm{LP}}, 2.0)$, while L-BFGS-B QAOA ($p = 6$) was run with a $45$-iteration limit ($663\text{--}819$ noiseless circuit evaluations per instance, mean $730$) at $N = 8$ and a $25$-iteration limit for $N = 6\text{--}14$. Although QZD \cite{Herman2023} was originally formulated with variational angle training (Table~\ref{tab:taxonomy}), here it was parameterized without outer-loop training via the QAOA-TQA linear ramp normalized by $\Delta E_{\mathrm{obj}}$ ($\gamma_{\ell} = 2.60s_{\ell}/\Delta E_{\mathrm{obj}}$, $\beta_{\ell} = -0.325(1 - s_{\ell})$) and $P_{\mathrm{QZD}}m_{\mathrm{sub}}$ projective $\Pi_{\mathcal{F}}$ checks without post-selection discard. Each Zeno slice retained $97.4\%\text{--}98.4\%$ feasible weight on average, which compounded over $40$ projections to $52.06\%$ survival on CSK and $34.60\%$ on QKP. Arithmetic-oracle GAS \cite{Gilliam2021} ($\approx 28N$ two-qubit gates per reflection) was evaluated in Fig.~\ref{fig:hw}(b), and full baseline equations are given in Appendix~\ref{app:baselines}.

The default parameters of Z-GRO were set to $M = 3$ ($M = 2$ in the scaling and noise sweeps), $r = 3$, $L_{\mathrm{epoch}} = 4$, $K_{\mathrm{filt}} = 6$, $\kappa = 0.35$, $k_{\mathrm{safe}} = 2.35$, $\alpha = 0.70$, and $\gamma_{\mathrm{rb}} = 0.50$ (see Table~\ref{tab:zgro_params} in Appendix~\ref{app:baselines}). Stationary fixed-point checks yielded $\|\mathcal{E}_{(3,k_{\mathrm{th}})}(\rho^{*}) - \rho^{*}\|_{F} < 10^{-15}$ for $k_{\mathrm{th}} \in \{1, 2\}$, and the $512$-trajectory Taylor--Walsh--Hadamard unraveling matched exact density-matrix evolution at $N = 8$ within $0.35\%\text{--}0.37\%$ root-mean-square error. Throughout Sec.~\ref{sec:evaluation} and Table~\ref{tab:results}, $\pm$ denotes one standard error of the mean ($\sigma / \sqrt{12}$, where $\sigma$ is the population standard deviation across the $12$ instances) for $P(z^{*})$ and one ordinary least-squares regression standard error of $b = 2^{\hat{\beta}}$ across $N \in \{6, 8, 10, 12, 14\}$ for $b$. All numerical simulations were executed in Python~3.11.2 using NumPy~2.1.3 and SciPy~1.15.2. The Python simulation and plotting routines were implemented with the assistance of Google Gemini (version 3.1 Pro) based on the author's mathematical models, and all outputs were verified against finite-size analytical checks.

\subsection{Convergence Dynamics and Five-Way Ablation Analysis}

As shown in Figs.~\ref{fig:conv_abl}(a) and \ref{fig:conv_abl}(b) and summarized in Table~\ref{tab:results}, the full adaptive Z-GRO scheme increases the mean optimal ground-state probability $P(z^{*})$ monotonically from $0.85\%$ to $13.70\% \pm 1.59\%$ over $T = 40$ macro-cycles for the CSK ensemble. For the QKP ensemble, $P(z^{*})$ rises from $1.00\%$ to $2.96\%$ at $t = 2$, decreases by $0.12$ percentage points to $2.84\%$ at the end of the coherent gauge launch ($t = 3$), and then increases monotonically across $t = 3\text{--}40$ to $12.13\% \pm 0.75\%$. The corresponding feasible-subspace fidelities are $99.99\%$ and $99.93\%$, respectively, and $0$ out of $24$ runs exhibit a threshold overshoot. Among the external baselines, fixed-schedule penalty-free QZD \cite{Herman2023} achieves $4.55\% \pm 0.73\%$ for CSK (with $52.06\%$ non-discarded feasible survival) and $2.85\% \pm 0.38\%$ for QKP (with $34.60\%$ survival). Conversely, variational L-BFGS-B QAOA achieves only $1.20\% \pm 0.13\%$ for CSK and $0.93\% \pm 0.03\%$ for QKP despite $663\text{--}819$ outer-loop circuit evaluations per instance ($45$ L-BFGS-B iterations). It is observed that FALQON ($1.20\% \pm 0.17\%$ and $0.67\% \pm 0.04\%$), QAOA-TQA ($0.80\% \pm 0.04\%$ and $0.75\% \pm 0.02\%$), and MITE ($0.61\% \pm 0.01\%$ and $0.60\% \pm 0.01\%$) also remain below $1.25\%$. In other words, Z-GRO outperforms fixed-schedule QZD by a factor of $3.0$ to $4.3$ and outperforms the penalized schemes by a factor of $11.4$ to $22.5$ in $P(z^{*})$.

Figures~\ref{fig:conv_abl}(c) and \ref{fig:conv_abl}(d) show the contribution of each component of Z-GRO by means of four controlled ablations. In Ablation~1, the threshold is fixed at $E_{\mathrm{ref}}^{(t)} \equiv E_{\mathrm{LP}}$ with a fixed $(r, k_{\mathrm{th}}) = (3, 1)$ ratchet gear. For CSK, the optimal probability drops from $13.70\%$ to $5.35\% \pm 2.09\%$ after $t = 3$ because the padded LP lower-bound gap (a mean gap of $-3.26$, or $-2.59$ before the $4\%$ safety margin) leaves the low-lying feasible excited states only weakly filtered. Conversely, for QKP, the initial McCormick relaxation is tight (a mean padded gap of $-1.08$, or $-0.68$ raw). Consequently, the static threshold achieves $11.38\% \pm 1.84\%$, which lies within one standard error of the adaptive result ($12.13\% \pm 0.75\%$). In Fig.~\ref{fig:conv_abl}(b), the static curve transiently exceeds the adaptive trajectory governed by the lower-tail safety margin in Eqs.~(\ref{eq:delta_theta})--(\ref{eq:cantelli_update}) for $t = 4\text{--}28$, whereas the adaptive scheme overtakes it for $t \ge 29$ with lower variance. In Ablation~2, the echoed $R_{\mathcal{F}}$ confinement is replaced with the penalized $H_{\mathrm{QUBO}}$ within the ancilla ratchet. As a result, the optimal probability collapses to $0.61\% \pm 0.12\%$ for CSK (with a feasibility of $63.38\%$) and $0.67\% \pm 0.05\%$ for QKP (with a feasibility of $50.88\%$). Note that this result is consistent with Proposition~\ref{prop:aliasing} and Fig.~\ref{fig:mech}(c), in which $\mathcal{G}_{\mathrm{kick}}(\mu)$ reaches $420.5$ at $\mu = 10$ and $38{,}438.5$ at $\mu = 100$. The latter value matches the analytical prediction of $38{,}502.6$ to within $0.17\%$. In Ablation~3, the dissipative ancilla ratchet is removed, while the echoed unitary gauge evolution is retained. The optimal probability reduces from $13.70\%$ to $9.02\% \pm 1.58\%$ for CSK and by a factor of $3.6$ (from $12.13\%$ to $3.38\% \pm 0.26\%$) for QKP as a result of reversible unitary dispersion. Finally, in Ablation~4, both the two-spin terms in $B_{\mathrm{corr}}$ ($\kappa = 0$) and the commutator gauge $G_{\mathrm{corr}}$ ($\eta_{\mathrm{rej}} = 0$, without the initial gauge launch) are omitted, and thus $K_{\mathrm{tunnel}}$ reduces to the single-spin mixer $\beta_{0}B_{1}$. The optimal probability decreases from $13.70\% \pm 1.59\%$ to $9.76\% \pm 1.35\%$ on CSK ($1.9$ combined standard errors), which is attributed to omitting the symmetry-breaking gauge launch and correlated two-spin transitions, whereas on QKP the single-spin $B_{1}$ generator achieves $11.60\% \pm 0.77\%$, within one standard error of the full scheme.

\begin{figure*}[!tp]
  \centering
  \includegraphics[width=\textwidth]{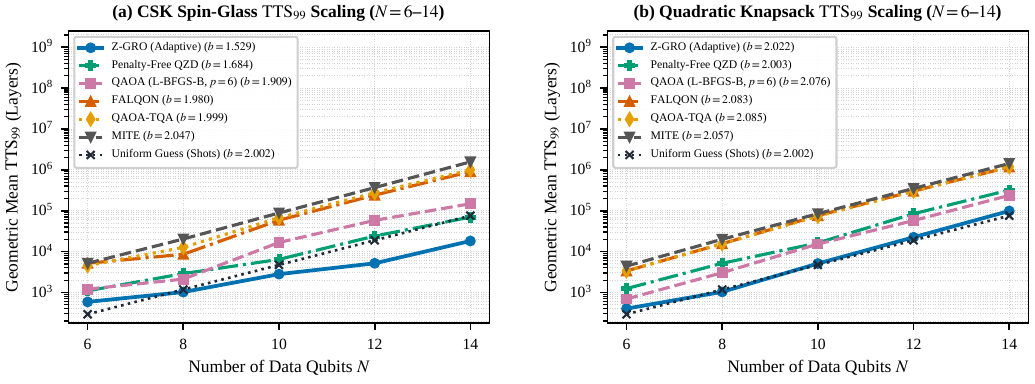}
  \caption{Exponential time-to-solution scaling $\mathrm{TTS}_{99}(N) \propto b^{N}$ in coherent circuit layers for $N \in \{6, 8, 10, 12, 14\}$ ($5$ instances per $N$ per ensemble, $T = 24$ layers or stages per quantum run, $p = 6$ for L-BFGS-B QAOA), evaluated using matrix-free quantum trajectory simulations. The dotted black curve plots the classical sample count $R_{99,\mathrm{unif}}(N)$ (in shots, one cost evaluation per sample, $b = 2.002$) of uniform random bitstring guessing. (a) CSK spin-glass ensemble ($0/25$ overshoots), for which Z-GRO achieves $b = 1.529 \pm 0.050$ compared to $1.684 \pm 0.039$ for fixed-schedule QZD \cite{Herman2023} and $1.909\text{--}2.047$ for the penalized schemes. (b) QKP ensemble ($3/25$ overshoots), in which all fixed-$T$ quantum schemes exhibit comparable bases near $2$ ($2.003\text{--}2.085$) while Z-GRO maintains the lowest $\mathrm{TTS}_{99}$ across all $N$ among the evaluated quantum schemes.}
  \label{fig:tts}
\end{figure*}

\begin{figure*}[!tp]
  \centering
  \includegraphics[width=\textwidth]{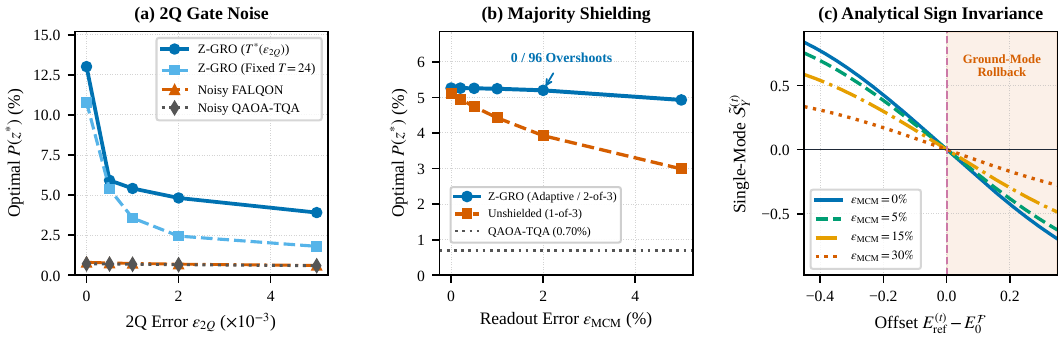}
  \caption{Hardware noise resilience for $N = 8$ ($M = 2$, dense $28$-edge coupling count, $16$ combined instances comprising $8$ CSK and $8$ QKP instances) and analytical quadrature sign-invariance schematic. (a) Mean ground-state probability $P(z^{*})$ under two-qubit depolarizing error $\epsilon_{2Q} \in [0, 5 \times 10^{-3}]$ ($\epsilon_{\mathrm{MCM}} = 0$) for Z-GRO at the post-hoc peak depth $T^{*}(\epsilon_{2Q}) \le 40$, Z-GRO at $T = 24$, and noisy baselines ($T = 24$). (b) Mean $P(z^{*})$ at $\epsilon_{2Q} = 5 \times 10^{-4}$ and a fixed depth of $T = 28$ under readout bit-flip error $\epsilon_{\mathrm{MCM}} \in [0\%, 5.0\%]$ for Z-GRO (adaptive gear at $\epsilon_{\mathrm{MCM}} \le 0.2\%$ and $(3, 2)$ majority shielding for $\epsilon_{\mathrm{MCM}} \ge 0.4\%$, $0/96$ overshoots), forced $(3, 1)$ unshielded ratcheting, and noisy QAOA-TQA. (c) Analytical single-eigenstate schematic plotting $(1 - 2\epsilon_{\mathrm{MCM}})\sin[-\tau(E_{\mathrm{ref}}^{(t)} - E_{0}^{\mathcal{F}})]$ ($\tau = 2.2$) versus $E_{\mathrm{ref}}^{(t)} - E_{0}^{\mathcal{F}}$ for pure $|z^{*}\rangle$, illustrating exact zero-crossing invariance for all $\epsilon_{\mathrm{MCM}} < 50\%$.}
  \label{fig:noise}
\end{figure*}

\subsection{Impact of System Size on Time-to-Solution Scaling}

To evaluate the asymptotic scalability of the proposed scheme across $5$ instances per $N$ ($80$ trajectories for $N \le 12$ and $48$ for $N = 14$), Fig.~\ref{fig:tts} shows the geometric mean of $\mathrm{TTS}_{99}(N) = T_{\mathrm{layers}} R_{99}(N)$ for $N \in \{6, 8, 10, 12, 14\}$, where $R_{99}(N) = \max(1, \ln(0.01)/\ln(1 - P(z^{*})))$ counts main optimization repetitions (excluding auxiliary Ehrenfest and Ramsey calibration shots, as well as QAOA L-BFGS-B outer-loop evaluations). Figure~\ref{fig:tts} also includes the classical shot-count reference $R_{99,\mathrm{unif}}(N)$ of uniform bitstring guessing ($b = 2.002 \pm 0.001$, $75{,}449$ samples at $N = 14$). On the CSK ensemble in Fig.~\ref{fig:tts}(a), Z-GRO records $0$ out of $25$ threshold overshoots and yields $b = 1.529 \pm 0.050$, with $\mathrm{TTS}_{99}$ equal to $586$, $1{,}021$, $2{,}803$, $5{,}165$, and $18{,}165$ coherent layers ($R_{99} = 757$ shots at $N = 14$). Fixed-schedule QZD \cite{Herman2023} yields $b = 1.684 \pm 0.039$ ($69{,}702$ stages and $R_{99} = 2{,}904$ shots at $N = 14$). Conversely, L-BFGS-B QAOA yields $b = 1.909 \pm 0.116$ ($148{,}015$ layers and $R_{99} = 24{,}669$ shots at $p = 6$), FALQON yields $b = 1.980 \pm 0.114$ ($904{,}418$ layers, $R_{99} = 37{,}684$ shots), QAOA-TQA yields $b = 1.999 \pm 0.051$ ($994{,}364$ layers, $R_{99} = 41{,}432$ shots), and MITE yields $b = 2.047 \pm 0.009$ ($1{,}567{,}213$ layers, $R_{99} = 65{,}301$ shots). Owing to penalty spectral compression, the single-run success probabilities of FALQON, QAOA-TQA, and MITE at $N = 14$ are only $1.2$ to $2.0$ times that of uniform bitstring guessing, and require more shots than ideal uniform sampling within $\mathcal{F}$ ($31{,}350$ samples for $|\mathcal{F}| = 6{,}808$). Overall, at $N = 14$ on CSK, Z-GRO achieves a $3.8$-fold speedup in $\mathrm{TTS}_{99}$ over fixed-schedule QZD and an $8.1$- to $86.3$-fold speedup over the penalized baselines, while requiring $99.7$ and $41.4$ times fewer shots than uniform bitstring and feasible-subspace sampling, respectively. Conversely, for the QKP ensemble in Fig.~\ref{fig:tts}(b) and Table~\ref{tab:results}, all quantum schemes at fixed $T = 24$ exhibit comparable scaling bases near $2$, namely $b = 2.003 \pm 0.038$ for QZD, $b = 2.022 \pm 0.064$ for Z-GRO, and $2.057\text{--}2.085$ for the penalized baselines. Across the $25$ QKP trajectory runs, $3$ instances exhibit undetected threshold overshoots (one at $N = 12$ across $t = 20\text{--}24$ by $0.007$, and two at $N = 14$ across $t = 20\text{--}24$ by $0.054$ and across $t = 8\text{--}24$ by up to $1.62$, or $6.8\%$ of $E_{\mathrm{ub}} - E_{0}^{\mathcal{F}}$), where $\tilde{\mathcal{S}}_{Y}^{(t)}$ remained positive because ground-state population in $w_{\mathrm{dark}}(z)$ was small relative to higher-energy feasible states. At $N = 14$, Z-GRO requires $R_{99} = 4{,}107$ shots ($18.4$ and $6.4$ times fewer than uniform bitstring guessing at $75{,}449$ and ideal feasible sampling at $26{,}451$), and reduces $\mathrm{TTS}_{99}$ to $98{,}568$ layers compared to $320{,}357$ stages for QZD ($3.3$ times) and $237{,}433\text{--}1{,}429{,}341$ layers for the penalized schemes ($2.4$ to $14.5$ times). Note that when weighted by $T = 24$ layers per shot on QKP at $N = 14$, the total layer count of $98{,}568$ exceeds the $75{,}449$ single-evaluation samples of uniform bitstring guessing.

\subsection{Impact of Two-Qubit Gate Errors and Mid-Circuit Readout Noise}

To evaluate hardware noise resilience on an equal footing, both sweeps in Figs.~\ref{fig:noise}(a) and \ref{fig:noise}(b) were conducted at $N = 8$ ($M = 2$) over $16$ combined instances ($8$ CSK and $8$ QKP instances) using an effective per-step local Pauli depolarizing probability $p_{\mathrm{dep}} = 1 - (1 - \epsilon_{2Q})^{n_{2Q}}$ parameterized by the dense coupling count $|E_{\mathrm{obj}}| = N(N - 1)/2 = 28$, while evaluating Ramsey quadratures from exact $(1 - 2\epsilon_{\mathrm{MCM}})$-scaled expectations without finite-shot sampling variance. Specifically, each layer of noisy QAOA-TQA and FALQON incurs $n_{2Q,\mathrm{base}} = |E_{\mathrm{obj}}| = 28$ two-qubit gates, each step of the three-step Z-GRO gauge launch ($t \in \{1, 2, 3\}$) incurs $n_{2Q,\mathrm{launch}} = 2|E_{\mathrm{obj}}| + 2N = 72$ gates, and each dissipative ratchet cycle ($t \ge 4$) incurs $n_{2Q}^{(t)} = r(2|E_{\mathrm{obj}}| + 4) + p_{\mathrm{kick}}^{(t)}(2|E_{\mathrm{obj}}| + 2N) = 60r + 72p_{\mathrm{kick}}^{(t)}$ gates across all $r \in \{2, 3\}$ executed ancilla checks and kicked fraction $p_{\mathrm{kick}}^{(t)}$. Figure~\ref{fig:noise}(a) sweeps $\epsilon_{2Q} \in \{0, 5 \times 10^{-4}, 10^{-3}, 2 \times 10^{-3}, 5 \times 10^{-3}\}$ up to $T = 40$ at $\epsilon_{\mathrm{MCM}} = 0$, where $T^{*}(\epsilon_{2Q})$ denotes the post-hoc maximum of the $16$-instance mean trajectory. At $\epsilon_{2Q} = 0$, this $M = 2$ benchmark rises from $5.99\%$ at the end of the gauge launch ($t = 3$) and $6.51\%$ at $t = 4$ to $13.01\%$ at $T^{*} = 40$ ($10.79\%$ at $T = 24$, $99.80\%$ feasibility). At $\epsilon_{2Q} = 5 \times 10^{-4}$, $P(z^{*})$ rises from $5.67\%$ ($t = 3$) to a peak of $5.92\%$ at $T^{*} = 4$ ($98.70\%$ feasibility) and retains $5.37\%$ at $T = 24$ ($99.06\%$ feasibility at $T = 40$), compared to $0.77\%$ for noisy FALQON and $0.70\%$ for noisy QAOA-TQA at $T = 24$. In this low-error regime ($\epsilon_{2Q} = 5 \times 10^{-4}$), conditional feed-forward limits depolarizing decay across $t = 4\text{--}24$ by bypassing the $72$-gate $V_{\mathcal{F}}$ block on retained dark trajectories, although the $60r$-gate ancilla checks still prevent net ratchet gain beyond $t = 4$. For $\epsilon_{2Q} \ge 10^{-3}$ under dense $28$-edge coupling, the cumulative check cost ($60r \in \{120, 180\}$ gates per cycle) exceeds single-cycle ratchet cooling, and thus the peak $P(z^{*})$ arises primarily from the three-step gauge launch ($t \le 3$) or the first cycle ($t = 4$) and declines thereafter. Specifically, Z-GRO records $(t = 3,\, t = 4,\, T = 24)$ probabilities of $(5.36\%, 5.41\%, 3.57\%)$ at $\epsilon_{2Q} = 10^{-3}$ ($T^{*} = 4$, $98.49\%$ feasibility, compared to $0.74\%$ for FALQON and $0.68\%$ for QAOA-TQA), $(4.82\%, 4.60\%, 2.44\%)$ at $2 \times 10^{-3}$ ($T^{*} = 3$, $81.52\%$ pre-reset feasibility rising to $98.14\%$ at $t = 4$), and $(3.61\%, 3.21\%, 1.80\%)$ at $5 \times 10^{-3}$ with peak $3.91\%$ at $T^{*} = 2$ ($78.99\%$ pre-reset feasibility rising to $97.53\%$ at $t = 4$, compared to $0.61\%$ for FALQON and $0.59\%$ for QAOA-TQA). Across all $80$ gate-error runs, $2$ runs on the same QKP instance exhibit late-cycle overshoots (by $0.016$ at $\epsilon_{2Q} = 10^{-3}$ across $t = 32\text{--}40$ and by $0.029$ at $2 \times 10^{-3}$ across $t = 24\text{--}40$, long after $T^{*} \in \{3, 4\}$).

Figure~\ref{fig:noise}(b) evaluates mid-circuit ancilla readout bit-flip errors for $\epsilon_{\mathrm{MCM}} \in \{0\%,\allowbreak 0.2\%,\allowbreak 0.5\%,\allowbreak 1.0\%,\allowbreak 2.0\%,\allowbreak 5.0\%\}$ across the same $16$ instances ($6 \times 16 = 96$ closed-loop runs) at $\epsilon_{2Q} = 5 \times 10^{-4}$ and a fixed depth of $T = 28$ macro-cycles (i.e., $25$ ratchet cycles after the gauge launch, rather than the post-hoc peak depth $T^{*} = 4$), where the default Z-GRO controller selects the adaptive gear for $\epsilon_{\mathrm{MCM}} \in \{0\%, 0.2\%\}$ and locks into the $(3, 2)$ majority-shielded gear for $\epsilon_{\mathrm{MCM}} \ge 0.4\%$ (Table~\ref{tab:zgro_params}). Over all $96$ runs, the controller in Eqs.~(\ref{eq:rollback})--(\ref{eq:cantelli_update}) records $0$ out of $96$ threshold overshoots ($E_{\mathrm{ref}}^{(t)} \le E_{0}^{\mathcal{F}}$ at every cycle $t$, with zero rollbacks triggered), consistent with Theorem~\ref{thm:control} and the analytical single-eigenstate schematic in Fig.~\ref{fig:noise}(c), which plots $(1 - 2\epsilon_{\mathrm{MCM}})\sin[-2.2(E_{\mathrm{ref}}^{(t)} - E_{0}^{\mathcal{F}})]$ for a pure ground eigenstate ($w_{\mathrm{dark}}(z^{*}) = 1$). To test active recovery under Eq.~(\ref{eq:rollback}) directly, an auxiliary experiment initialized $|z^{*}\rangle\langle z^{*}|$ across all $24$ $N = 8$ instances at an overshot reference $E_{\mathrm{ref}}^{(0)} = E_{0}^{\mathcal{F}} + 0.35(E_{\mathrm{ub}} - E_{0}^{\mathcal{F}})$ over $20$ noisy cycles ($\epsilon_{2Q} = 5 \times 10^{-4}$, $\epsilon_{\mathrm{MCM}} = 5.0\%$, $r = 3$ checks per cycle). Enabling Eq.~(\ref{eq:rollback}) detected $\tilde{\mathcal{S}}_{Y}^{(t)} < 0$ in all $24$ runs and retained $P(z^{*}) = 27.78\% \pm 0.31\%$, whereas disabling rollback collapsed $P(z^{*})$ to $0.51\% \pm 0.08\%$. Note that the remaining decay from the initial $100\%$ under rollback arises mainly from the depolarizing gate channel over the $20$ cycles. At $\epsilon_{\mathrm{MCM}} = 0\%$ in Fig.~\ref{fig:noise}(b), default Z-GRO and the forced $(3, 1)$ unshielded variant start at $5.26\%$ and $5.10\%$, respectively. As $\epsilon_{\mathrm{MCM}}$ increases to $5.0\%$, the unshielded $(3, 1)$ ratchet drops by $41.4\%$ (from $5.10\%$ to $2.99\%$) due to linear $\mathcal{O}(\epsilon_{\mathrm{MCM}})$ false dark-state kicks, whereas default Z-GRO under $(3, 2)$ majority shielding drops by only $6.3\%$ (from $5.26\%$ to $4.93\%$, with $5.24\%$ at $\epsilon_{\mathrm{MCM}} = 1.0\%$), consistent with Eq.~(\ref{eq:majority_shield}).

Finally, Fig.~\ref{fig:hw}(b) shows the entangling gate counts of bounded-degree ($d = 3$) instances at $T = 12$ for $N = 20$ to $50$ qubits. It is observed that the shallow Z-GRO compilation requires only $599$ to $1{,}484$ active $R_{ZZ}$ gates on trapped-ion hardware, and $343$ to $563$ parallel Rydberg pulses with $3$ epoch camera readouts on neutral-atom arrays. These values represent a reduction by a factor of $45$ to $119$ relative to reversible $28N$-gate adder comparators ($26{,}880$ to $67{,}200$ two-qubit gates).

\section{Conclusion}
\label{sec:conclusion}

This paper has proposed an accurate, penalty-free, outer-loop-free, readout-error-tolerant, scalable, and cost-effective dynamic quantum circuit scheme, designated as Z-GRO, for solving constrained combinatorial optimization problems. In the proposed scheme, reversible feasibility phase operators interleaved within symmetric dynamical-decoupling echoes confine the state evolution to the feasible subspace without penalty multipliers, while fused commutator gauge tunneling resolves the stall of symmetric feedback. Directional ground-state convergence is driven by a binomial ancilla spectral ratchet based on zero-two-qubit-gate virtual-$Z$ threshold offsets and conditional mid-circuit gate bypassing, and the threshold is regulated by a sign-invariant Ramsey quadrature controller. The performance of the proposed scheme has been evaluated by performing a series of exact density-matrix and matrix-free quantum trajectory simulations using frustrated CSK and QKP ensembles. Overall, the simulation results have shown that Z-GRO achieves a single-check discrimination gain of $38{,}438.5$ at $\mu = 100$, improves the optimal ground-state probability by a factor of $3.0$ to $4.3$ over fixed-schedule QZD, and improves the probability by a factor of $11.4$ to $22.5$ over variational QAOA, FALQON, QAOA-TQA, and MITE. Furthermore, it has been shown that Z-GRO achieves an exponential $\mathrm{TTS}_{99}$ scaling base of $1.529 \pm 0.050$ on the CSK ensemble for $N = 6$ to $14$ qubits, compared to $1.684 \pm 0.039$ for QZD and $1.909\text{--}2.047$ for the penalized schemes. On the QKP ensemble, in which the exponential bases remain comparable near $2$ at $T = 24$, Z-GRO reduces $\mathrm{TTS}_{99}$ at $N = 14$ by a factor of $3.3$ relative to QZD and by a factor of $2.4$ to $14.5$ relative to the penalized schemes. In addition, the platform-native compilation requires $45$ to $119$ times fewer entangling operations than arithmetic comparator oracles for $N = 20$ to $50$ qubits. Moreover, it has been shown that Z-GRO maintains an optimal probability higher than that of noisy FALQON by a factor of $2.9$ to $7.0$ at $T = 24$ for two-qubit gate error rates between $5 \times 10^{-4}$ and $5 \times 10^{-3}$, although the ancilla check cost under dense coupling prevents a net ratchet gain beyond $t = 4$. Finally, it has been shown that the $(3, 2)$ majority gear limits the loss in the optimal probability to $6.3\%$ at a mid-circuit readout bit-flip rate of $5.0\%$, and that the threshold controller records $0/24$ overshoots on the $N = 8$ noiseless benchmarks, $0/25$ on CSK scaling, and $0/96$ across the readout-noise sweep (with $3/25$ overshoots on QKP scaling and $2/80$ late-cycle overshoots in the gate-error sweep). The future work will investigate adaptive circuit-depth schedules $T(N)$ for larger system sizes, experimental demonstrations of the proposed scheme on zoned neutral-atom and QCCD trapped-ion processors, and extensions of the scheme to mixed-integer dynamic scheduling problems.

\section*{Competing Interests}
The author is the inventor on a pending U.S. Provisional Patent Application (No. 64/202,093) covering the dynamic quantum circuit architectures, hardware-native constraint compilation methods, and closed-loop threshold control protocols described in this work.

\section*{Data Availability}
The data that support the findings of this article are not publicly available. The data are available from the authors upon reasonable request.

\appendix
\renewcommand{\thesection}{\Alph{section}}
\renewcommand{\theequation}{\Alph{section}\arabic{equation}}
\setcounter{equation}{0}

\section{Proofs of Proposition~\ref{prop:aliasing} and Theorem~\ref{thm:zeno}}
\label{app:aliasing}

\begin{proof}[Proof of Proposition~\ref{prop:aliasing}]
Consider Problem (P2) with a threshold of $E_{\mathrm{ref}} = E_{0}^{\mathcal{F}}$, where $E_{\mathrm{QUBO}}(z) = E_{\mathrm{obj}}(z) + \mu E_{\mathrm{pen}}(z)$. On $\mathcal{F}$, it holds that $E_{\mathrm{pen}}(z) = 0$. Over all of the bitstrings, $\max_{z \in \{0,1\}^{N}}(E_{\mathrm{QUBO}}(z) - E_{0}^{\mathcal{F}}) \le \Delta E_{\mathrm{obj}} + \mu P_{\max}$, where the equality is attained when the maximum objective and penalty spans coincide. The phase $\Phi(z) = \tau(E_{\mathrm{QUBO}}(z) - E_{0}^{\mathcal{F}})/2$ must remain within $[0, \pi/2]$ to prevent the aliasing of the high-energy infeasible states, and thus the maximum evolution time is $\tau_{\mathrm{pen}} = \pi / (\Delta E_{\mathrm{obj}} + \mu P_{\max})$. Conversely, when the evolution is confined to $\mathcal{F}$, the non-aliasing evolution time is $\tau_{\mathrm{Z\text{-}GRO}} = \pi / \Delta E_{\mathrm{obj}}$. For the first feasible excited state $E_{1}^{\mathcal{F}} = E_{0}^{\mathcal{F}} + \delta E_{1}$ with $\delta E_{1} \ll \Delta E_{\mathrm{obj}}$, the Taylor expansion $\sin^{2}(x) = x^{2} - x^{4}/3 + \mathcal{O}(x^{6})$ yields
\begin{equation}
  \begin{aligned}
    \frac{\sin^{2}(\tau_{\mathrm{Z\text{-}GRO}}\delta E_{1}/2)}{\sin^{2}(\tau_{\mathrm{pen}}\delta E_{1}/2)} &= \frac{(\pi \delta E_{1} / 2\Delta E_{\mathrm{obj}})^{2}}{(\pi \delta E_{1} / 2(\Delta E_{\mathrm{obj}} + \mu P_{\max}))^{2}} \\
    &\quad \times \bigl[1 + \mathcal{O}(\delta E_{1}^{2})\bigr],
  \end{aligned}
\end{equation}
which simplifies directly to Eq.~(\ref{eq:kick_gain}).
\end{proof}

\begin{proof}[Proof of Theorem~\ref{thm:zeno}]
Let $\delta t = \theta / (2M)$ and consider a single echo pair $W(\delta t) = R_{\mathcal{F}}e^{-i\delta t(K_{\parallel} + K_{\perp})}R_{\mathcal{F}}e^{-i\delta t(K_{\parallel} + K_{\perp})}$. Since $R_{\mathcal{F}}^{2} = I$, $R_{\mathcal{F}}K_{\parallel}R_{\mathcal{F}} = K_{\parallel}$, and $R_{\mathcal{F}}K_{\perp}R_{\mathcal{F}} = -K_{\perp}$, it follows that $R_{\mathcal{F}}e^{-i\delta t(K_{\parallel}+K_{\perp})}R_{\mathcal{F}} = e^{-i\delta t(K_{\parallel}-K_{\perp})}$. The application of the Baker--Campbell--Hausdorff expansion to $e^{-i\delta t(K_{\parallel}-K_{\perp})}e^{-i\delta t(K_{\parallel}+K_{\perp})}$ cancels the linear $-i\delta t K_{\perp}$ terms identically, i.e.,
\begin{equation}
  W(\delta t) = \exp\!\left(-i2\delta t K_{\parallel} + \delta t^{2}[K_{\perp}, K_{\parallel}] + \mathcal{O}(\delta t^{3}\|K\|_{2}^{3})\right).
\end{equation}
Since $[K_{\perp}, K_{\parallel}]$ maps $\mathrm{Ran}(\Pi_{\mathcal{F}})$ to $\mathrm{Ran}(\Pi_{\mathcal{F}^{\perp}})$, the off-block transition norm of a single pair is bounded by $\delta t^{2}\|[K_{\parallel}, K_{\perp}]\|_{2} + \mathcal{O}(\delta t^{3}\|K\|_{2}^{3})$. The telescopic unitary error bound $\| \Pi_{\mathcal{F}^{\perp}} W(\delta t)^{M}|\psi\rangle \|_{2} \le M \| \Pi_{\mathcal{F}^{\perp}} W(\delta t) \Pi_{\mathcal{F}} \|_{2}$ with $\delta t = \theta / (2M)$ then gives $M(\theta/2M)^{2}\|[K_{\parallel}, K_{\perp}]\|_{2} = \theta^{2}\|[K_{\parallel}, K_{\perp}]\|_{2}/(4M)$, which establishes Eq.~(\ref{eq:leak_bound}).
\end{proof}

\setcounter{equation}{0}
\section{Proof of Proposition~\ref{prop:gauge}}
\label{app:gauge}

\begin{proof}
Let $G_{\mathrm{corr},\parallel} = \Pi_{\mathcal{F}}G_{\mathrm{corr}}\Pi_{\mathcal{F}} + \Pi_{\mathcal{F}^{\perp}}G_{\mathrm{corr}}\Pi_{\mathcal{F}^{\perp}}$ denote the block-diagonal component of $G_{\mathrm{corr}}$. For any integer $M \ge 1$, applying the product rule at $\eta = 0$ to $U_{\mathrm{Z\text{-}Echo}}(G_{\mathrm{corr}}, \eta, M) = [R_{\mathcal{F}}e^{-i\eta G_{\mathrm{corr}}/(2M)}R_{\mathcal{F}}e^{-i\eta G_{\mathrm{corr}}/(2M)}]^{M}$ with $R_{\mathcal{F}}^{2} = I$ gives $\frac{d}{d\eta}U_{\mathrm{Z\text{-}Echo}}(G_{\mathrm{corr}}, \eta, M)\big|_{\eta=0} = -\frac{i}{2}(R_{\mathcal{F}}G_{\mathrm{corr}}R_{\mathcal{F}} + G_{\mathrm{corr}}) = -iG_{\mathrm{corr},\parallel}$ identically without finite-$M$ error. Differentiating $E(\eta) = \langle \psi_{\mathcal{F}} | U_{\mathrm{Z\text{-}Echo}}^{\dagger}(G_{\mathrm{corr}}, \eta, M) H_{\mathrm{obj}} U_{\mathrm{Z\text{-}Echo}}(G_{\mathrm{corr}}, \eta, M) | \psi_{\mathcal{F}} \rangle$ at $\eta = 0$ therefore yields $E'(0) = \langle \psi_{\mathcal{F}} | i[G_{\mathrm{corr},\parallel}, H_{\mathrm{obj}}] | \psi_{\mathcal{F}} \rangle$. For the uniform feasible state $|\psi_{\mathcal{F}}\rangle = |\mathcal{F}|^{-1/2}\sum_{z\in\mathcal{F}}|z\rangle$, note that $(B_{\mathrm{corr}})_{z,z'} \ge 0$ for all $z, z' \in \mathcal{F}$, while $H_{\mathrm{obj}}$ is diagonal in the computational basis. Specifically, $(G_{\mathrm{corr},\parallel})_{z,z'} =\allowbreak i(B_{\mathrm{corr}})_{z,z'}\allowbreak(E_{\mathrm{obj}}(z') - E_{\mathrm{obj}}(z))$. Expanding the commutator expectation on $|\psi_{\mathcal{F}}\rangle$ yields
\begin{align}
  E'(0) &= -\frac{1}{|\mathcal{F}|}\sum_{z,z'\in\mathcal{F}} (B_{\mathrm{corr}})_{z,z'}\bigl(E_{\mathrm{obj}}(z') - E_{\mathrm{obj}}(z)\bigr)^{2},
\end{align}
which establishes Eq.~(\ref{eq:gauge_slope}) and is strictly negative whenever $(B_{\mathrm{corr}})_{z,z'} > 0$ connects at least one pair of feasible states with $E_{\mathrm{obj}}(z) \neq E_{\mathrm{obj}}(z')$. Fitting the quadratic Taylor approximation $E(\eta) \approx E(0) + g_{1}\eta + \frac{1}{2}g_{2}\eta^{2}$ at $\{0, +\delta, -\delta\}$ via central finite differences gives $\eta^{*} = \mathrm{clip}(-g_{1}/g_{2}, -\eta_{\max}, +\eta_{\max})$ when $g_{2} > \epsilon_{\mathrm{curv}}$ and the fixed-sign step $\eta^{*} = -\eta_{\mathrm{fb}}\,\mathrm{sgn}(g_{1})$ when $g_{2} \le \epsilon_{\mathrm{curv}}$.
\end{proof}

\setcounter{equation}{0}
\section{Proof of Theorem~\ref{thm:ratchet}}
\label{app:ratchet}

\begin{proof}
First, the trace preservation of Eq.~(\ref{eq:binomial_cptp}) is verified. Since $V_{\mathcal{F}}^{\dagger}V_{\mathcal{F}} = I$ and $K_{r,s}^{\dagger}K_{r,s} = \cos^{2(r-s)}(\Phi_{t})\sin^{2s}(\Phi_{t})$, summing over $s = 0, \dots, r$ via the binomial theorem yields $\sum_{s=0}^{r}\binom{r}{s}\cos^{2(r-s)}(\Phi_{t})\sin^{2s}(\Phi_{t}) = (\cos^{2}\Phi_{t} + \sin^{2}\Phi_{t})^{r} = I$. Next, when $E_{\mathrm{ref}}^{(t)} = E_{0}^{\mathcal{F}}$, every optimal feasible state $z^{*} \in \mathcal{F}^{*}$ satisfies $\Phi_{t}(z^{*}) = 0$, and thus $\cos\Phi_{t}(z^{*}) = 1$ and $\sin\Phi_{t}(z^{*}) = 0$. Consequently, $K_{r,0}|z^{*}\rangle = |z^{*}\rangle$ and $K_{r,s}|z^{*}\rangle = 0$ for all $s \ge 1$. Since $k_{\mathrm{th}} \ge 1$, only the retained $s = 0$ term acts non-trivially on $\rho^{*}$ supported on $\mathcal{F}^{*}$, which yields $\mathcal{E}_{(r,k_{\mathrm{th}})}(\rho^{*}) = K_{r,0}\rho^{*}K_{r,0}^{\dagger} = \rho^{*}$. Conversely, for any excited feasible state $z \in \mathcal{F} \setminus \mathcal{F}^{*}$, it holds that $\Phi_{t}(z) \in [\Phi_{\min}, \pi/2]$ with $\Phi_{\min} = \tau_{t}\delta E_{1}/2 > 0$, and thus the per-cycle rejection probability satisfies $p_{\mathrm{rej}}(z) = \sum_{s=k_{\mathrm{th}}}^{r}\binom{r}{s}\sin^{2s}\Phi_{t}(z)\cos^{2(r-s)}\Phi_{t}(z) \ge p_{\mathrm{rej}}(\Phi_{\min}) > 0$. By ergodicity, a fraction of at least $\gamma_{\min} = \min_{z\in\mathcal{F}\setminus\mathcal{F}^{*}}\sum_{z^{*}\in\mathcal{F}^{*}}|\langle z^{*}|V_{\mathcal{F}}|z\rangle|^{2} > 0$ of each rejected excited population is transferred irreversibly into the absorbing manifold $\mathcal{F}^{*}$ per macro-cycle.
\end{proof}

\setcounter{equation}{0}
\section{Proofs of Theorem~\ref{thm:control} and Proposition~\ref{prop:ms}}
\label{app:control}
\label{app:hardware}

\begin{proof}[Proof of Theorem~\ref{thm:control}]
Consider an ideal ancilla bit $b \in \{0, 1\}$ with the observable $Z_{a} = 1 - 2b \in \{+1, -1\}$. The bit is passed through a symmetric readout flip channel, which produces the noisy bit $\tilde{b} = b \oplus \xi$, where $\Pr(\xi = 1) = \epsilon_{\mathrm{MCM}}$. The conditional expectation of $1 - 2\tilde{b}$ given $b$ is $(1 - \epsilon_{\mathrm{MCM}})(1 - 2b) + \epsilon_{\mathrm{MCM}}(2b - 1) = (1 - 2\epsilon_{\mathrm{MCM}})(1 - 2b)$. In the $Y$-quadrature Ramsey check on the computational state $|z\rangle$, the ideal ancilla state prior to the $Z$-basis measurement has $\mathbb{E}[1 - 2b \mid z] = \sin(\tau_{t}(E_{\mathrm{obj}}(z) - E_{\mathrm{ref}}^{(t)}))$, which yields Eq.~(\ref{eq:sign_invariance}) after averaging over $w_{\mathrm{dark}}(z)$. When $E_{\mathrm{ref}}^{(t)} \le E_{0}^{\mathcal{F}}$, it holds that $0 \le \tau_{t}(E_{\mathrm{obj}}(z) - E_{\mathrm{ref}}^{(t)}) \le \pi$ for all $z \in \mathcal{F}$, and thus $\sin(\tau_{t}(E_{\mathrm{obj}}(z) - E_{\mathrm{ref}}^{(t)})) \ge 0$ for every term. Since $1 - 2\epsilon_{\mathrm{MCM}} > 0$ for all $\epsilon_{\mathrm{MCM}} < 0.5$, it follows that $\mathbb{E}[\tilde{\mathcal{S}}_{Y}^{(t)}] \ge 0$. Furthermore, for a ground dark state (i.e., $\Phi_{t}(z^{*}) = 0$), the independent readout flips over $r = 3$ checks follow the distribution $\mathrm{Binomial}(3, \epsilon_{\mathrm{MCM}})$. As a result, $\Pr(S \ge 2) = \binom{3}{2}\epsilon_{\mathrm{MCM}}^{2}(1 - \epsilon_{\mathrm{MCM}}) + \binom{3}{3}\epsilon_{\mathrm{MCM}}^{3}$, which matches Eq.~(\ref{eq:majority_shield}).
\end{proof}

\begin{proof}[Proof of Proposition~\ref{prop:ms}]
Conjugating the global MS operator $\exp[-i\frac{\pi}{8}(\sum_{i=1}^{N}X_{i})^{2}]$ by $R_{y}(\pi/2)^{\otimes N}$ transforms $X_{i} \mapsto Z_{i} = 1 - 2z_{i}$, and thus produces $\exp[-i\frac{\pi}{8}(N - 2W(z))^{2}]$. Substituting $W(z) = C_{0} + \Delta$ expands the phase in the exponent as
\begin{equation}
  \begin{aligned}
    -\frac{\pi}{8}(N - 2C_{0} - 2\Delta)^{2} &= -\frac{\pi}{8}(N - 2C_{0})^{2} \\
    &\quad + \frac{\pi}{2}(N - 2C_{0})\Delta - \frac{\pi}{2}\Delta^{2}.
  \end{aligned}
\end{equation}
Based on the integer identity $\Delta^{2} = \Delta(\Delta - 1) + \Delta$, the sum of the quadratic and linear terms in $\Delta$ is equal to $-\frac{\pi}{2}\Delta(\Delta - 1) + \frac{\pi}{2}(N - 2C_{0} - 1)\Delta$. The single-qubit rotations $\bigotimes_{i=1}^{N}R_{z}(-\pi(N - 2C_{0} - 1)/2)$ cancel the linear $\sum_{i}z_{i}$ term up to a global constant phase $\gamma_{0}$. The remaining phase is $\exp[-i\frac{\pi}{2}\Delta(\Delta - 1)] = (-1)^{\Delta(\Delta - 1)/2}$, which is equal to $+1$ for $\Delta \in \{0, 1\}$ and $-1$ for $\Delta \in \{-2, -1, +2, +3\}$.
\end{proof}

\setcounter{equation}{0}
\section{Z-GRO Implementation Parameters and Baseline Formulations}
\label{app:baselines}

Table~\ref{tab:zgro_params} lists all implementation parameters and control rules of Z-GRO used across the numerical simulations. In particular, after solving the linear programs in Eq.~(\ref{eq:p3_lp}) for the raw bounds $(E_{\mathrm{LP}}^{\mathrm{raw}}, E_{\mathrm{ub}}^{\mathrm{raw}})$ with span $\Delta_{\mathrm{LP}} = E_{\mathrm{ub}}^{\mathrm{raw}} - E_{\mathrm{LP}}^{\mathrm{raw}}$, a $\pm 4\%$ spectral safety margin $E_{\mathrm{LP}} = E_{\mathrm{LP}}^{\mathrm{raw}} - 0.04\Delta_{\mathrm{LP}}$ and $E_{\mathrm{ub}} = E_{\mathrm{ub}}^{\mathrm{raw}} + 0.04\Delta_{\mathrm{LP}}$ is applied so that the phase $\tau_{t}(E_{\mathrm{obj}}(z) - E_{\mathrm{ref}}^{(t)})$ stays strictly inside $(0, \pi)$ near the upper spectral edge. At $N = 8$, the mean raw lower-bound gaps $E_{\mathrm{LP}}^{\mathrm{raw}} - E_{0}^{\mathcal{F}}$ are $-2.59$ on CSK and $-0.68$ on QKP, which become $-3.26$ and $-1.08$ after the $4\%$ safety margin.

\begin{table*}[!tp]
  \centering
  \caption{Complete reproducibility parameters and closed-loop control rules of Z-GRO in the simulation suites.}
  \label{tab:zgro_params}
  \footnotesize
  \begin{tabularx}{\textwidth}{@{}llX@{}}
    \toprule
    \textbf{Module / Mechanism} & \textbf{Symbol} & \textbf{Simulation Setting and Implementation Rule} \\
    \midrule
    McCormick LP safety margin & $E_{\mathrm{LP}},\, E_{\mathrm{ub}}$ & $E_{\mathrm{LP}} = E_{\mathrm{LP}}^{\mathrm{raw}} - 0.04\Delta_{\mathrm{LP}}$, $E_{\mathrm{ub}} = E_{\mathrm{ub}}^{\mathrm{raw}} + 0.04\Delta_{\mathrm{LP}}$ ($\Delta_{\mathrm{LP}} = E_{\mathrm{ub}}^{\mathrm{raw}} - E_{\mathrm{LP}}^{\mathrm{raw}}$) \\
    Feasible state initialization & $n_{\mathrm{init}}$ & $4$ projective $\Pi_{\mathcal{F}}$ reload rounds from $|+\rangle^{\otimes N}$, transferring residual infeasible weight onto $|z_{\mathrm{seed}}\rangle$ \\
    Echo slice pairs & $M$ & $M = 3$ in the $N = 8$ noiseless benchmarks and ablations ($M = 2$ in the scaling and noise suites) \\
    Correlated driver & $\kappa,\, |\mathcal{P}_{\mathrm{top}\text{-}N}|$ & $\kappa = 0.35$ on up to $N$ objective edges $(u,v)$ with largest $|J_{uv}|$ \\
    Coherent gauge launch & $n_{\mathrm{launch}}$ & $3$ echoed $G_{\mathrm{corr}}$ descent steps with unitary diagonal phase rotation $\exp[-i(0.12/\max(\sigma_{E}, 0.5))H_{\mathrm{obj}}]$ before steps $2, 3$, with $\sigma_{E} = \mathrm{std}_{z}(E_{\mathrm{obj}}(z))$ over $\{0, 1\}^{N}$ \\
    Ehrenfest step calibration & $\delta,\, \eta_{\max},\, \epsilon_{\mathrm{curv}}$ & $\delta = 0.012$, $\eta_{\max} = 0.16$, $\epsilon_{\mathrm{curv}} = 0.25$ (fixed fallback $\eta^{*} = -0.06\,\mathrm{sgn}(g_{1})$ when $g_{2} \le 0.25$) \\
    Fused tunneling generator & $\beta_{0},\, \eta_{\mathrm{rej}}$ & $K_{\mathrm{tunnel}} = \beta_{0}B_{\mathrm{corr}} + \eta_{\mathrm{rej}}G_{\mathrm{corr}}$ with $\beta_{0} = 0.58$ and $\eta_{\mathrm{rej}} = 0.75\eta_{\mathrm{calib}}^{*}$ \\
    Epoch & $L_{\mathrm{epoch}},\, K_{\mathrm{filt}}$ & $L_{\mathrm{epoch}} = 4$ macro-cycles between $E_{\mathrm{ref}}^{(t)}$ updates and leaky-state resets onto $V_{\mathcal{F}}|z_{\mathrm{seed}}\rangle$, $K_{\mathrm{filt}} = 6$ \\
    Threshold controller & $k_{\mathrm{safe}},\, \alpha,\, \gamma_{\mathrm{rb}},\, \epsilon_{\mathrm{tol}}$ & $k_{\mathrm{safe}} = 2.35$, $\alpha = 0.70$, $\gamma_{\mathrm{rb}} = 0.50$, $\epsilon_{\mathrm{tol}} = 0$ \\
    Dynamic ratchet gear $(r, k_{\mathrm{th}})$ & $q_{\mathrm{hi}},\, q_{\mathrm{lo}}$ & $(3, 2)$ if $q_{\mathrm{dark}}^{(t)} > 0.065$ or $\epsilon_{\mathrm{MCM}} \ge 0.4\%$, $(2, 1)$ if $0.032 < q_{\mathrm{dark}}^{(t)} \le 0.065$, and $(3, 1)$ if $q_{\mathrm{dark}}^{(t)} \le 0.032$ \\
    \bottomrule
  \end{tabularx}
\end{table*}

To evaluate all penalized baselines on the identical $N$-qubit Hilbert space ($\dim = 2^{N}$) without expanding the register by binary slack qubits (which favors the baselines by avoiding slack-space dilution), the constraint violations in Problem~(P2) are encoded directly as exact diagonal hinge-squared operators. For the CSK ensemble with $L = \lfloor 0.25N \rfloor$ and $U = \lfloor 0.50N \rfloor$, the penalty function is
\begin{equation}
  \begin{aligned}
    E_{\mathrm{pen}}^{\mathrm{CSK}}(z) &= \Bigl(\max\bigl(0,\, L - W(z)\bigr) + \max\bigl(0,\, W(z) - U\bigr) \\
    &\quad + z_{1}z_{2} + z_{3}z_{4}\Bigr)^{2},
  \end{aligned}
  \label{eq:epen_csk}
\end{equation}
where $W(z) = \sum_{i=1}^{N}z_{i}$. For the QKP ensemble with $L = \lfloor 0.22\sum_{i=1}^{N}w_{i} \rfloor$, $U = \lfloor 0.45\sum_{i=1}^{N}w_{i} \rfloor$, and mean weight $\bar{w} = N^{-1}\sum_{i=1}^{N}w_{i}$, the normalized penalty function is
\begin{equation}
  \begin{aligned}
    E_{\mathrm{pen}}^{\mathrm{QKP}}(z) &= \frac{1}{\bar{w}^{2}}\Bigl(\max\bigl(0,\, L - {\textstyle\sum_{i=1}^{N}w_{i}z_{i}}\bigr) \\
    &\quad + \max\bigl(0,\, {\textstyle\sum_{i=1}^{N}w_{i}z_{i}} - U\bigr)\Bigr)^{2}.
  \end{aligned}
  \label{eq:epen_qkp}
\end{equation}
Across all system sizes $N \in \{6, \dots, 14\}$, the four external penalized schemes set $E_{\mathrm{QUBO}}(z) = E_{\mathrm{obj}}(z) + \mu E_{\mathrm{pen}}(z)$ with the instance-calibrated multiplier $\mu = 2\max(E_{\mathrm{ub}} - E_{\mathrm{LP}}, 2.0)$ and spectral span $\Delta E_{\mathrm{QUBO}} = \max_{z}E_{\mathrm{QUBO}}(z) - \min_{z}E_{\mathrm{QUBO}}(z)$ (whereas Ablation~2 sets $\mu_{\mathrm{abl}} = \max(12.0, 2\|H_{\mathrm{obj}}\|_{1})$ to test severe QUBO spectral compression inside the ancilla ratchet).

Starting from $|+\rangle^{\otimes N}$, QAOA-TQA applies $T$ alternating layers $e^{-i\beta_{\ell}B_{1}}e^{-i\gamma_{\ell}H_{\mathrm{QUBO}}}$ ($\ell = 1, \dots, T$) along the linear ramp $s_{\ell} = \ell/(T + 1)$ with $\Delta t = 0.65$, $\gamma_{\ell} = 4\Delta t\,s_{\ell}/\Delta E_{\mathrm{QUBO}}$, and $\beta_{\ell} = -0.5\Delta t(1 - s_{\ell})$. Variational QAOA ($p = 6$) initializes $(\gamma_{\ell}, \beta_{\ell})$ on linear ramps over $[0.4, 2.4]/\Delta E_{\mathrm{QUBO}}$ and $[-0.45, -0.08]$ and optimizes $\langle H_{\mathrm{QUBO}}\rangle$ via SciPy L-BFGS-B with an iteration cap of $45$ iterations ($663\text{--}819$ circuit evaluations per instance via forward-difference gradients requiring $13$ circuit evaluations per step at $N = 8$) and $25$ iterations for $N = 6\text{--}14$. Fixed-schedule QZD \cite{Herman2023} initializes $|\psi_{\mathcal{F}}\rangle$ and applies $P_{\mathrm{QZD}}$ outer layers ($P_{\mathrm{QZD}} = 10, m_{\mathrm{sub}} = 4$ at $N = 8$, and $P_{\mathrm{QZD}} = 8, m_{\mathrm{sub}} = 3$ for $N = 6\text{--}14$) with $s_{\ell} = \ell/(P_{\mathrm{QZD}} + 1)$, $\gamma_{\ell} = 2.60s_{\ell}/\Delta E_{\mathrm{obj}}$, and $m_{\mathrm{sub}}$ interleaved steps $[\Pi_{\mathcal{F}}e^{-i(\beta_{\ell}/m_{\mathrm{sub}})B_{1}}]^{m_{\mathrm{sub}}}$ where $\beta_{\ell} = -0.325(1 - s_{\ell})$. FALQON \cite{Magann2022} breaks initial symmetry via $|\psi_{0}\rangle = e^{-i\Delta t_{\mathrm{cost}}H_{\mathrm{QUBO}}}|+\rangle^{\otimes N}$ ($\Delta t_{\mathrm{cost}} = 1.5/\Delta E_{\mathrm{QUBO}}$) and updates $\beta_{t} =\allowbreak \mathrm{clip}(-1.8\langle i[B_{1}, H_{\mathrm{QUBO}}]\rangle_{t-1}/\allowbreak\mathcal{N}_{\mathrm{comm}},\allowbreak -0.6,\allowbreak +0.6)$ with $\mathcal{N}_{\mathrm{comm}} = \max(\|i[B_{1}, H_{\mathrm{QUBO}}]\|_{2}, 1)$ at $N = 8$ and $\max(0.35N\Delta E_{\mathrm{QUBO}}, 1)$ in the matrix-free suite. Finally, MITE \cite{Mao2023} applies weak-measurement Kraus operators $M_{0}(z) = \cos[\epsilon_{\mathrm{MITE}}(E_{\mathrm{QUBO}}(z) - E_{\mathrm{mid}}) + \pi/4]$ and $M_{1}(z) = \sin[\epsilon_{\mathrm{MITE}}(E_{\mathrm{QUBO}}(z) - E_{\mathrm{mid}}) + \pi/4]$ centered at $E_{\mathrm{mid}} = \frac{1}{2}(\max_{z}E_{\mathrm{QUBO}} + \min_{z}E_{\mathrm{QUBO}})$ with $\epsilon_{\mathrm{MITE}} = 0.48\pi/\Delta E_{\mathrm{QUBO}}$, followed by conditional mixing $e^{-i(0.47)B_{1}}$ upon $m = 1$ ($80$ quantum trajectories per instance for $N \in \{6, 8, 10, 12\}$ and $48$ trajectories for $N = 14$).

\end{document}